\pdfoutput=1
\documentclass[prl,twocolumn,superscriptaddress]{revtex4-2}

\usepackage{amsmath,amsthm,mathtools}
\usepackage{amssymb,amsfonts}
\usepackage{upgreek}
\usepackage{graphicx,floatrow,float}
\usepackage{multirow}
\usepackage[italicdiff]{physics}
\usepackage{comment}
\usepackage[normalem]{ulem}
\usepackage{mathrsfs}
\usepackage{bbm}
\usepackage{dsfont}
\usepackage{chemformula}

\usepackage{hyperref}
\usepackage{color}
\definecolor{supcol}{RGB}{10,50,180}
\definecolor{eqcol}{RGB}{220,10,100}
\hypersetup{
    colorlinks,
    citecolor=supcol,
    linkcolor=eqcol,
    urlcolor=supcol
}

\makeatletter
\renewcommand*\l@section{\@dottedtocline{1}{0em}{2.4em}}
\renewcommand*\l@subsection{\@dottedtocline{2}{1.6em}{2.4em}}
\makeatother

\allowdisplaybreaks

\begin{document}

\makeatletter
\let\HSE@addcontentsline\addcontentsline
\renewcommand{\addcontentsline}[3]{}
\makeatother

\title{Experimental Investigation of Tunable-Order Hilbert-Space Ergodicity}

\author{Zou-Wei Pan}
\affiliation{School of Physics, Institute of Fundamental and Transdisciplinary Research, Institute of Quantum Sensing, State Key Laboratory of Ocean Sensing, and Zhejiang Key Laboratory of R$\&$D and Application of Cutting-edge Scientific Instruments, Zhejiang University, Hangzhou, 310058, China}

\author{Wenquan Liu}
\email{liuwenquan@zju.edu.cn}
\affiliation{School of Physics, Institute of Fundamental and Transdisciplinary Research, Institute of Quantum Sensing, State Key Laboratory of Ocean Sensing, and Zhejiang Key Laboratory of R$\&$D and Application of Cutting-edge Scientific Instruments, Zhejiang University, Hangzhou, 310058, China}

\author{Xing Rong}
\email{xrong@ustc.edu.cn}
\affiliation{School of Physics, Institute of Fundamental and Transdisciplinary Research, Institute of Quantum Sensing, State Key Laboratory of Ocean Sensing, and Zhejiang Key Laboratory of R$\&$D and Application of Cutting-edge Scientific Instruments, Zhejiang University, Hangzhou, 310058, China}

\begin{abstract}
Hilbert-space ergodicity (HSE) provides a new framework for studying thermalization in driven quantum systems, complementing the eigenstate thermalization hypothesis, which is restricted to static systems.
This ergodicity is hierarchical: by quantifying how randomly the dynamics explores the Hilbert space, one obtains a family of levels termed $k$-HSE.
While HSE has been observed at the lowest and highest levels, finite-order HSE dynamics remains largely unexplored due to the difficulty of constructing such drives.
Here, we explore this intermediate regime and uncover its distinctive physics.
We first propose and prove that a family of $m$-tone drives on qubits realizes $k$-HSE up to $k = 2m{-}3$, with drive parameters determined at $O(k)$ cost.
Using a single nitrogen-vacancy center in diamond, we verify this design by showing that a 3-tone drive realizes 3-HSE, with fourth-order statistics depending on the initial state.
Further in-depth theoretical analysis shows that this initial-state dependence is generic across drives, demonstrating the possibility of recovering the initial state from higher-order statistics even when the dynamics is ergodic.
Our work broadens the study of quantum ergodicity and reveals intriguing physics within its hierarchy.
\end{abstract}
\maketitle

Hilbert-space ergodicity (HSE), a recently proposed form of quantum ergodicity, offers a new framework for studying quantum thermalization that complements the eigenstate thermalization hypothesis~\cite{Berry1977,Deutsch1991,Srednicki1994,Rigol2008,DAlessio2016,Gogolin2016,Ueda2020}, and has attracted considerable research interest~\cite{PilatowskyCameo2023,PilatowskyCameo2024,PilatowskyCameo2025,Logaric2025}.
This notion of ergodicity is characterized by the statistical similarity between the temporal ensemble, generated by the time evolution of a quantum system, and a prescribed maximum-entropy (Haar) ensemble.
Specifically, by evaluating the trace distance between the $k$-th moments of these ensembles, one determines whether the dynamics exhibits $k$-HSE; interestingly, increasing $k$ allows one to assess progressively stronger forms of ergodicity.
The emergence of $k$-HSE thus reveals that quantum ergodicity is more intricate than its classical counterpart~\cite{Boltzmann1896,Birkhoff1931,LebowitzPenrose1973,CornfeldFominSinai1982,TodaKuboSaito}, potentially yielding richer physics.
These ergodicity conditions form a strict hierarchy: $k'$-HSE implies $k$-HSE whenever $k' > k$, with complete HSE (CHSE)---where the condition holds for all $k$---lying at the apex.

\begin{figure}[!b]
\centering
\includegraphics[width=1\columnwidth]{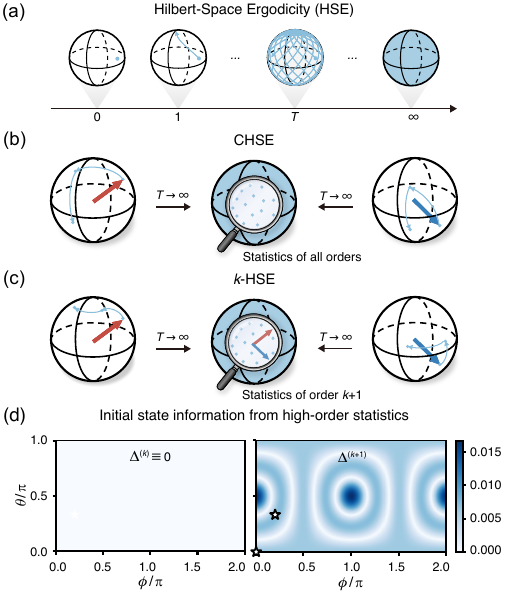}
\caption{
\textbf{Hilbert-space ergodicity.}
(a) Analogous to classical ergodicity, HSE describes the property that quantum dynamics uniformly explores the Hilbert space in the long-time limit.
(b)--(c) Comparison between HSE at infinite order (b, CHSE) and at finite order (c, $k$-HSE). A distinctive feature of $k$-HSE dynamics is that initial-state information survives in the statistics of order $k{+}1$ and higher.
(d) A qubit example, with the model specified in the text. Here $(\theta,\phi)$ denote the polar and azimuthal angles of the initial state.
}
\label{fig:concept}
\end{figure}

While signatures of HSE have been found in a Rydberg atom quantum simulator~\cite{Shaw2025} and the hierarchy of HSE was observed in a spin system~\cite{Liu2026}, the finite-order ergodic dynamics remains largely unexplored, despite its potential to reveal new phenomena.
Consider a quantum dynamics exhibiting $k$-HSE but not $(k{+}1)$-HSE: the temporal ensemble then converges to the Haar ensemble for statistical moments up to order $k$, so that its properties at these orders are fully captured by the Haar ensemble.
For higher-order moments, however, the temporal ensemble deviates from the Haar ensemble and should therefore contain additional information.
Yet what this information encodes, and what unique physics such finite-order quantum ergodicity may harbor, remain open questions.
In addition to being a nascent frontier topic, this problem has remained out of reach because constructing tunable ergodic dynamics is highly nontrivial.
To date, only one construction has been proposed~\cite{PilatowskyCameo2024}, whose complexity grows exponentially with $k$, hindering practical exploration of finite-order HSE physics.

In this Letter, we systematically investigate this problem from both theoretical and experimental perspectives, making three key contributions. 
First, focusing on qubits, we design a family of $m$-tone quasiperiodic drives and prove that they can realize $k$-HSE up to $k=2m{-}3$. 
Notably, our design is computationally efficient: with drive parameters determined at $O(k)$ cost, the drive generates a temporal ensemble forming a quantum state $k$-design, and hence realizes $k$-HSE.
By contrast, prior proposals either are
  tailored to a specific $k$ or require computational resources that scale exponentially with $k$~\cite{PilatowskyCameo2024,Liu2026}.
Second, using a single nitrogen-vacancy (NV) center in diamond, we experimentally verify this design by demonstrating that a 3-tone drive realizes 3-HSE but not 4-HSE.
Additionally, we find that the fourth-order trace distance varies with the initial state.
Finally, our in-depth theoretical analysis shows that the fourth-order trace distance generally depends on the initial state, and that this dependence is not specific to the 3-tone drive but extends to drives with other numbers of tones.
This reveals a counterintuitive feature of finite-order ergodic dynamics: although initial-state information is lost up to order $k$, it survives in the $(k{+}1)$-th and higher-order statistics (see Fig.~\ref{fig:concept}).
This property sets finite-order HSE apart from both classical ergodic dynamics---where initial-state information cannot be recovered from late-time observables---and quantum CHSE dynamics, where such information is lost at all orders of statistics.

\textit{Hilbert-space ergodicity and $m$-tone drives.}---We first recall the definition of HSE, and then introduce a family of $m$-tone quasiperiodic drives, showing how they realize $k$-HSE up to $k = 2m{-}3$.
HSE concerns the statistical similarity between the temporal ensemble $\{|\psi(t)\rangle\}_{t\ge0}$, generated by the unitary dynamics $U(t)$ of the system from an initial state $|\psi(0)\rangle$, and the Haar ensemble obtained by uniformly sampling the $d$-dimensional Hilbert space.
Mathematically, this similarity is quantified by the trace distance $\Delta_T^{(k)} = \frac{1}{2} \bigl\| \rho_{T}^{(k)} - \rho_{\text{Haar}}^{(k)} \bigr\|_1$, where $\rho_T^{(k)} = \frac{1}{T} \int_0^T \mathrm{d}t \, (|\psi(t)\rangle \langle \psi(t)|)^{\otimes k}$ is the $k$-th moment of the temporal ensemble, and $\rho_{\text{Haar}}^{(k)} = \int_{\mathrm{SU}(d)} \mathrm{d}V \, |V\rangle \langle V|^{\otimes k}$ is that of the Haar ensemble, with $|V\rangle = V|0\rangle$ and $\mathrm{d}V$ the Haar measure on $\mathrm{SU}(d)$~\cite{NielsenChuang2010,BookWatrous2018}.
If $\Delta_T^{(k)} \to 0$ as $T \to \infty$ (or falls below a sufficiently small threshold at large but finite $T$) for every $|\psi(0)\rangle$, the evolution $U(t)$ is said to satisfy $k$-HSE.
CHSE corresponds to the limit where $k$-HSE holds for all $k$.

\begin{figure}[t]
\centering
\includegraphics[width=1\columnwidth]{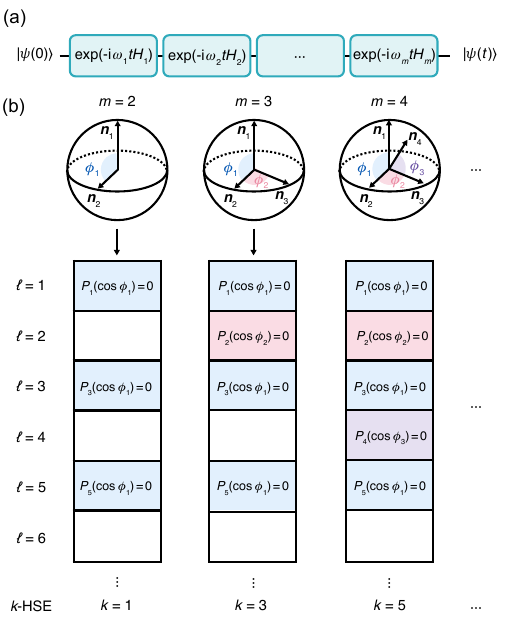}
\caption{
\textbf{Realizing tunable HSE with $m$-tone drives.}
(a) Design of the drive: a sequence of $m$ rotations about distinct axes $H_j = \boldsymbol{n}_j \cdot \boldsymbol{\sigma}$, with rationally independent frequencies $\{\omega_j\}$.
(b) Progressively realizing higher-order HSE by adding rotation axes and tuning their angles.
The rows labeled by $\ell$ represent the $\ell$-th HSE equation in Eq.~\eqref{eq:k-HSE condition}.
Each equation is fulfilled by adjusting $\phi_j = \angle(\boldsymbol{n}_j, \boldsymbol{n}_{j+1})$ with satisfied conditions indicated by colored cells.
The number of consecutively colored cells counted from the top determines the $k$-HSE order of the $m$-tone drive.
Notably, setting $\phi_1 = \pi/2$ satisfies all odd rows. Each even row is then filled by adding one additional rotation axis at a time, so that an $m$-tone drive can achieve $k$-HSE up to $k = 2m{-}3$.
}
\label{fig:construction}
\end{figure}

From the observable viewpoint, HSE imposes constraints on the temporal fluctuations of an arbitrary observable $O$.
The $k$-th temporal moment of $O$ reads
\begin{equation}
    \overline{O^{k}} = \lim_{T\to\infty}\frac{1}{T} \int_0^T \mathrm{d}t\, \langle\psi(t)|O|\psi(t)\rangle^{k} = \mathrm{tr}[O^{\otimes k}\rho_{\infty}^{(k)}].
    \label{eq1}
\end{equation}
Under $k$-HSE, this moment converges to the Haar average $\mathrm{tr}[O^{\otimes k}\rho_{\text{Haar}}^{(k)}]$, independent of the initial state.
At $k=1$, Eq.~\eqref{eq1} fixes the time average $\overline{O}$ of the quantum expectation value $\langle\psi(t)|O|\psi(t)\rangle$, but leaves its temporal variance $\overline{O^{2}}-\overline{O}^{2}$ unspecified.
For $k \ge 2$, it additionally fixes this variance and other fluctuations up to order $k$, while fluctuations of order $k{+}1$ and beyond remain unconstrained.
In the CHSE limit, moments of all orders are fixed to their Haar values, eliminating any structured fluctuation and forcing the quantum dynamics to explore the Hilbert space in a random manner, without any preferred pattern~\cite{Liu2026}.

Following previous studies~\cite{Martin2017,Dumitrescu2018,Else2020,Long2022,He2025}, we design each $m$-tone drive as a sequence of successive single-axis rotations at $m$ rationally independent frequencies~\cite{Weyl1916}:
\begin{equation}
U(t) = e^{-i\omega_m t H_m}\cdots e^{-i\omega_2 t H_2} e^{-i\omega_1 t H_1},
\label{eq:m-tone}
\end{equation}
where $H_j = \boldsymbol{n}_j \cdot \boldsymbol{\sigma}$, with $\{\omega_j\}$ the rotation frequencies and $\{\boldsymbol{n}_j\}$ unit vectors specifying the rotation axes.
A schematic of these drives is shown in Fig.~\ref{fig:construction}(a). In the Supplemental Material (SM)~\cite{SM}, we prove that the drive realizes $k$-HSE when the following $k$ equations hold:
\begin{equation}
    \prod_{j=1}^{m-1} P_\ell(\cos\phi_j) = 0, \quad \ell = 1, 2, \ldots, k,
    \label{eq:k-HSE condition}
\end{equation}
where $\phi_j = \angle(\boldsymbol{n}_j, \boldsymbol{n}_{j+1})$ is the angle between successive rotation axes, and $P_\ell$ is the Legendre polynomial of degree $\ell$.
Figure~\ref{fig:construction}(b) illustrates how these conditions are progressively satisfied by adding driving frequencies and tuning the corresponding angles.
The left-hand side of the $\ell$-th equation in Eq.~\eqref{eq:k-HSE condition} is a product of $P_\ell(\cos\phi_j)$, so the equation holds whenever $\cos\phi_j$ is a zero of $P_\ell$ for at least one $j$, as indicated by the colored cells in Fig.~\ref{fig:construction}(b).
Since Legendre polynomials of odd degree are odd functions, setting $\phi_1 = \pi/2$ automatically satisfies all equations with odd $\ell$, corresponding to the dark blue cells in Fig.~\ref{fig:construction}(b).
Consequently, 1-HSE is realized by 2-tone drives.
Adding one further driving frequency at a time and solving the associated even-order conditions for $\phi_j$ then raises the ergodicity order up to $k = 2m{-}3$, consistent with the expectation~\cite{PilatowskyCameo2024} that higher-order ergodicity requires more rationally independent driving frequencies.
Therefore, our scheme generates quantum dynamics with tunable $k$-HSE at a computational cost of $O(k)$~\cite{Bogaert2014}.

Equipped with these designs, we now explicitly construct a finite-order ergodic quantum dynamics and explore its unique physics.
Here, we use a 3-tone drive for illustration; additional examples are provided in the SM~\cite{SM}.
The drive frequencies are set to $\omega_1=1$, $\omega_2=(1+\sqrt{2})/2$, and $\omega_3=(1+\sqrt{3})/2$ to ensure their rational independence.
The corresponding rotation axes are chosen as $\boldsymbol{n}_1=\boldsymbol{e}_x$, $\boldsymbol{n}_2=(\boldsymbol{e}_x+\sqrt{2}\boldsymbol{e}_y)/\sqrt{3}$, and $\boldsymbol{n}_3=(-\sqrt{2}\boldsymbol{e}_x+\boldsymbol{e}_y)/\sqrt{3}$, such that
  $\phi_2=\pi/2$, while $\cos\phi_1=1/\sqrt{3}$ ensures $P_2(\cos\phi_1)=0$.
Note that permuting the indices of $\phi_j$ in Eq.~\eqref{eq:k-HSE condition} does not alter the corresponding HSE property~\cite{SM}.
To evaluate the ergodicity, we sample the temporal ensemble at stroboscopic times $t\in\mathbb{N}$.
The resulting trace distances are plotted in Fig.~\ref{fig:concept}(d): $\Delta_\infty^{(k)}$ vanishes for $k \leq 3$ irrespective of the initial state (left panel), while $\Delta_\infty^{(4)}$ remains nonzero (right panel), confirming that the dynamics satisfies $3$-HSE but not $4$-HSE.
Furthermore, the distinct dependence of $\Delta_\infty^{(4)}$ on the parameters of the initial state indicates that initial-state information survives in higher-order statistics.
Such survival is impossible for classical ergodic dynamics or for quantum CHSE dynamics, and hence constitutes a unique feature of finite-order HSE dynamics.

\begin{figure}[!t]
\centering
\includegraphics[width=\textwidth]{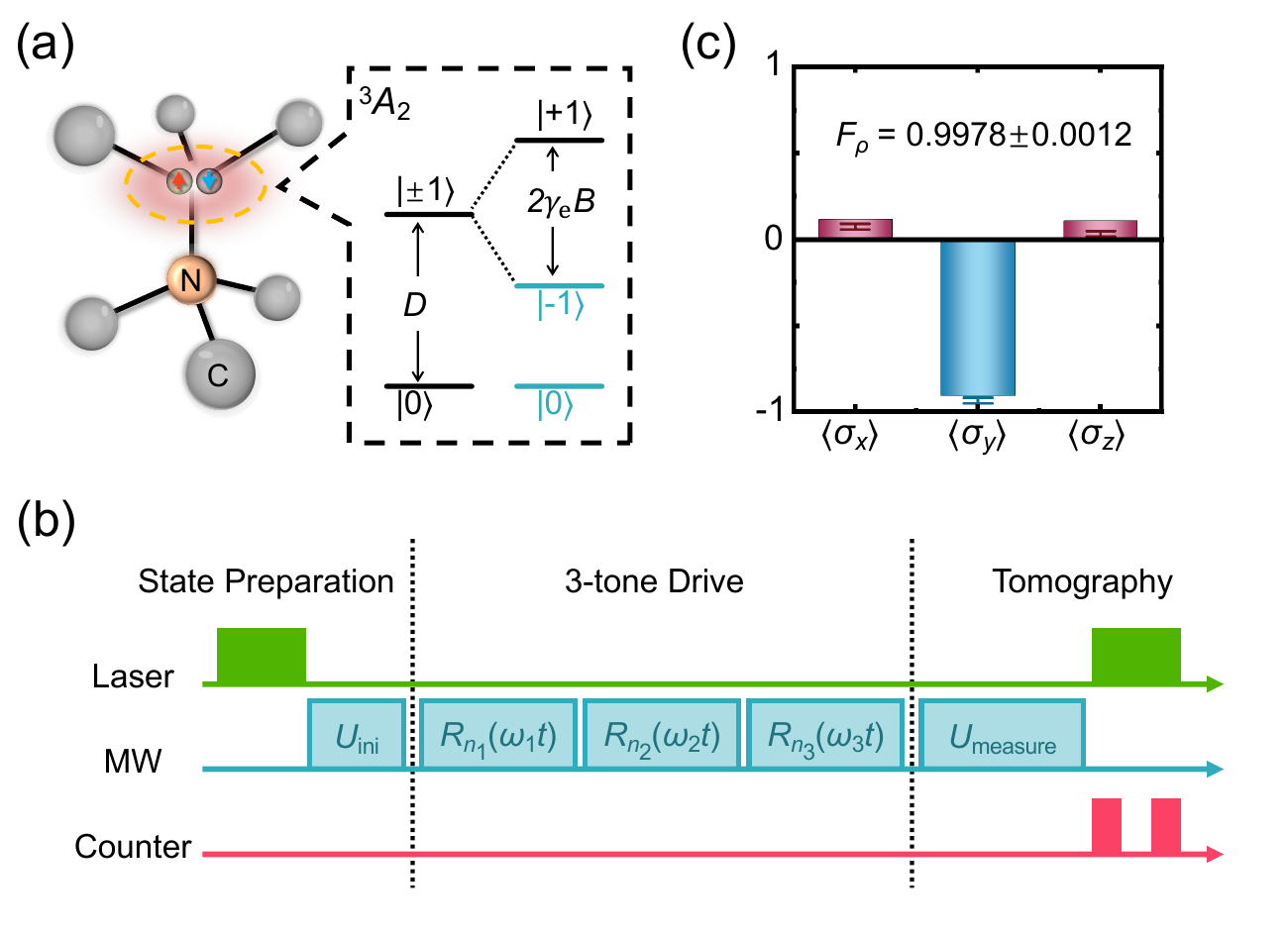}
\caption{
\textbf{NV center system and experimental scheme.}
(a) Diamond lattice and energy levels of the NV center. The electron-spin states $\ket{m_s=0}\equiv\ket{0}$ and $\ket{m_s={-}1}\equiv\ket{1}$ are identified as the experimental qubit.
(b) Experimental pulse sequence, consisting of state preparation, execution of the 3-tone drive, and quantum state tomography. Measurement operations $U_{\mathrm{measure}} \in \{I, R_X(\pi), R_X({\pm}\pi/2), R_Y({\pm}\pi/2)\}$ are used to read out the full density matrix.
(c) Illustrative tomography result for $|\psi(0)\rangle = |0\rangle$ and $t=100$. Dots with error bars represent experimental data, and gray bars correspond to theoretical predictions.
}
\label{fig:experiment}
\end{figure}

\textit{Experimental verification.}---We experimentally implement the 3-tone drive in a single NV center, verifying that the dynamics exhibits 3-HSE and that initial-state information can be detected in the fourth-order statistics.
An NV center is a point defect in the diamond lattice, consisting of a substitutional nitrogen atom paired with an adjacent vacancy [Fig.~\ref{fig:experiment}(a)]~\cite{Gali2008,Maze2011,Doherty2013}.
When negatively charged, the $\mathrm{NV}^{-}$ center hosts a triplet ground state $^3A_2$, whose sublevels $|m_s = 0\rangle$ and $|m_s = {\pm}1\rangle$ are split by a zero-field splitting $D=2.87$~GHz.
We additionally apply a static magnetic field of $B\sim503$~G along the NV axis, which further lifts the degeneracy of the $|m_s = {\pm}1\rangle$ sublevels, yielding a splitting of $2\gamma_e B$ and facilitating the polarization of the neighboring nitrogen nuclear spin~\cite{Jacques2009,Busaite2020}.
We drive the $|m_s = 0\rangle \leftrightarrow |m_s = {-}1\rangle$ transition with resonant microwave pulses, thereby defining an effective qubit with $|0\rangle \equiv |m_s = 0\rangle$ and $|1\rangle \equiv |m_s = {-}1\rangle$; meanwhile, the far-detuned $|m_s = {+}1\rangle$ state can be safely ignored~\cite{deLange2010,Choi2017}.
Details of the diamond sample and NV properties are provided in the SM~\cite{SM}.

The pulse sequence for realizing and detecting the quantum dynamics is illustrated in Fig.~\ref{fig:experiment}(b), which consists of three stages.
First, a $3$~$\upmu$s 532 nm laser pulse followed by a microwave operation $U_{\mathrm{init}}$ is applied to polarize the spin and prepare the initial state $|\psi(0)\rangle = U_{\mathrm{init}}|0\rangle$.
Then, three additional microwave pulses are performed to realize the target $3$-tone drive. In repeated experiments, the pulse durations are varied to generate the state $|\psi(t)\rangle$ at different times, thereby forming the temporal ensemble.
Finally, the readout of $|\psi(t)\rangle$ is achieved by applying a microwave operation $U_{\mathrm{measure}}$ followed by a laser pulse to collect the spin-state-dependent fluorescence. By varying $U_{\mathrm{measure}}$ and repeating the pulse sequence, quantum state tomography of $|\psi(t)\rangle$ is performed.
Figure~\ref{fig:experiment}(c) exemplifies the reconstructed state $|\psi(t)\rangle$ at $t=100$, starting from the initial state $|\psi(0)\rangle = |0\rangle$.
The experimental results are in excellent agreement with theoretical predictions, yielding a quantum state fidelity of $0.9978 \pm 0.0012$ and demonstrating the high-precision generation and detection of the quantum dynamics.

\begin{figure}[t]
\centering
\includegraphics[width=\columnwidth]{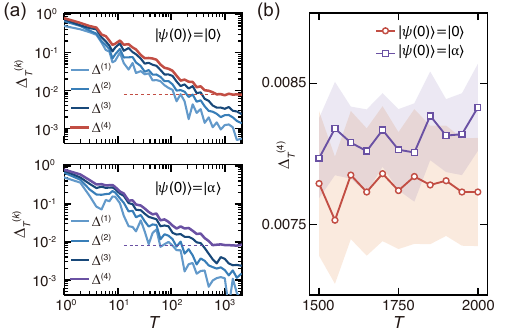}
\caption{
\textbf{Experimental results for the 3-tone drive.}
(a) Trace distances $\Delta_T^{(k)}$ up to the fourth order for initial states $|\psi(0)\rangle=|0\rangle$ (upper panel) and $|\psi(0)\rangle=|\alpha\rangle\equiv\cos\frac{\pi}{6}|0\rangle + e^{i\pi/5}\sin\frac{\pi}{6}|1\rangle$ (lower panel).
Both axes are displayed on a logarithmic scale, and the dashed line indicates the plateau value of $\Delta_T^{(4)}$.
(b) Zoomed-in plot of $\Delta_T^{(4)}$.
Shaded regions represent experimental uncertainties.
}
\label{fig:results}
\end{figure}

Figure~\ref{fig:results} presents the experimental results.
For a given initial state $\ket{\psi(0)}$, we evolved the system under the 3-tone drive for 2000 steps and recorded the temporal ensemble $\{|\psi(t)\rangle\}_{t=0}^{1999}$.
Using this discrete ensemble to approximate the continuous-time temporal ensemble, we computed the moments $\rho_T^{(k)} = (1/T) \sum_{t=0}^{T-1} (|\psi(t)\rangle\langle\psi(t)|)^{\otimes k}$ and the trace distance $\Delta_T^{(k)}$ up to the fourth order; this approximation is valid provided that $\omega_1, \ldots, \omega_m$ together with $2\pi$ are rationally independent.
The upper panel of Fig.~\ref{fig:results}(a) shows the results for $\ket{\psi(0)} = |0\rangle$.
The trace distances for the first three orders decrease overall with increasing $T$, showing no sign of saturation, whereas $\Delta_T^{(4)}$ initially decays but then plateaus at a nonzero value after approximately $1000$ steps.
To demonstrate that this behavior is intrinsic to the drive and independent of the initial state, we repeated the measurement with another initial state, $|\psi(0)\rangle = \cos\frac{\pi}{6}|0\rangle + e^{i\pi/5}\sin\frac{\pi}{6}|1\rangle$.
Similar behavior is observed in the lower panel.
These results unambiguously demonstrate that the 3-tone drive indeed realizes $k$-HSE up to $k=3$, consistent with our theoretical design.

After verifying the HSE property of the 3-tone drive, we now turn to the initial-state information encoded in the higher-order statistics of the temporal ensemble.
The monotonic decay of $\Delta_T^{(k)}$ for $k \in \{1,2,3\}$, which is independent of the initial state, indicates that the information about $|\psi(0)\rangle$ is completely erased from the statistics up to the third order.
Furthermore, one can infer from the non-4-HSE property of the dynamics that $\Delta_T^{(4)}$ does not converge to zero but instead plateaus at a nonzero value.
However, upon zooming in on the late-time window of the dynamics, as shown in Fig.~\ref{fig:results}(b), we find that the plateau value actually differs between the two initial states.
Prompted by this observation, we further evaluate $\Delta_\infty^{(4)}$ for other initial states, producing the color map shown in Fig.~\ref{fig:concept}(d), which displays a distinctive pattern encoding the initial-state information.

Finally, to understand this feature and explore the underlying physics, we perform an analytical study and extend the result to other drives.
By parametrizing the initial state as $\rho = (I + \boldsymbol{r}\cdot\boldsymbol{\sigma})/2$ and computing the fourth-order statistics, we find $\Delta_\infty^{(4)} = |P_4(\cos\gamma)|/60$, where $\gamma$ is the angle between the Bloch vector $\boldsymbol{r}$ and the first rotation axis $\boldsymbol{n}_1$.
This expression explains several features of Fig.~\ref{fig:concept}(d).
First, the figure exhibits three maxima located at $\theta = \pi/2$ with $\phi = 0, \pi, 2\pi$. 
These maxima originate from the fact that, for these initial states, the first rotation $e^{-i\omega_1 t H_1}$ contributes only a global phase, effectively reducing the 3-tone drive to a 2-tone drive and thereby resulting in a relatively large $\Delta_\infty^{(4)}$.
Second, the pattern exhibits clear symmetry and periodicity.
These features stem from the fact that the information retained in $\Delta_\infty^{(4)}$ depends solely on the angle between the Bloch vector of the initial state and the first rotation axis ($x$ axis here), making Fig.~\ref{fig:concept}(d) symmetric about $\theta=\pi/2$ and periodic in $\phi$.
To demonstrate that these properties are intrinsic to finite-order HSE dynamics rather than specific to the 3-tone drive, we investigated them for drives with other numbers of tones.
The results are presented in the SM~\cite{SM}, where similar behavior is observed, consistent with our conjecture: for dynamics that is $k$-HSE but not $(k{+}1)$-HSE, the initial-state information, although hidden from the statistics up to order $k$, emerges in the statistics of order $k{+}1$ and higher.

\textit{Conclusion.}---In this work, we conducted a systematic investigation of finite-order Hilbert-space ergodic quantum dynamics.
To address the theoretical challenge of constructing tunable HSE dynamics, we proposed and proved that a family of $m$-tone quasiperiodic drives can realize $k$-HSE up to order $k = 2m - 3$, with the drive parameters determined at a computational cost of $O(k)$.
Using a single NV center in diamond, we experimentally confirmed this design by demonstrating that a 3-tone drive realizes 3-HSE but not 4-HSE.
In addition, we found that the fourth-order trace distance depends on the initial state.
Furthermore, our theoretical analysis shows that this initial-state dependence is not specific to the 3-tone drive but constitutes a distinct property of finite-order HSE.
This property therefore sets finite-order HSE apart from both classical ergodic dynamics and quantum CHSE dynamics, establishing it as a distinctive dynamical regime in the study of ergodicity and thermalization.

Building on this work, two questions immediately arise and can be explored.
The first concerns the initial state: since the statistics of order $k{+}1$ and higher encode measurable signatures of $|\psi(0)\rangle$, a natural question is whether $|\psi(0)\rangle$ can be fully reconstructed from them.
An affirmative answer would turn this hidden fingerprint into a practical resource for quantum information processing and metrology.
The second concerns higher-dimensional systems. 
Since the current construction relies on the representation theory of SU(2), extending it to arbitrary $d$-dimensional systems would establish tunable ergodicity as a universal feature of driven quantum dynamics.

More broadly, advances in quantum technologies have made it feasible to measure and record time-resolved quantum trajectories of individual quantum states, going beyond the time-averaged or spatially averaged observables accessible in classical measurements.
This capability opens the door to re-examining the relation between ergodicity and thermalization, long regarded as intimately linked in classical systems.
For example, recent works have proposed a generalized maximum entropy principle~\cite{Mark2024,MokScrooge2026} that unifies the ergodic dynamics of closed quantum systems with the quantum deep thermalization of open subsystems (obtained by projective measurements on the rest of the system)~\cite{Ho2022,Ippoliti2022,Choi2023,Cotler2023,Liu2024,Chang2025,Chakraborty2025}.
We anticipate that future advances will extend the $m$-tone drive to arbitrary $d$-dimensional systems, enabling exploration of the interplay between ergodicity and thermalization in quantum many-body systems.

\begin{acknowledgments}
{\it Acknowledgments.}---
The authors thank Wen Wei Ho, Yue Fu and Yunhan Wang for insightful discussions. 
This work is supported by the National Natural Science Foundation of China (Grant Nos. 12475027, T2388102, 12261160569), the Major Science and Technology Projects of Zhejiang Province (Grant No. 2025C02027), the Innovation Program for Quantum Science and Technology (Grant No. 2021ZD0302200), and the Fundamental Research Funds for the Central Universities (Grant No. 226-2024-00011). 
\end{acknowledgments}

\bibliography{kHSE}

\begin{thebibliography}{45}%
\makeatletter
\providecommand \@ifxundefined [1]{%
 \@ifx{#1\undefined}
}%
\providecommand \@ifnum [1]{%
 \ifnum #1\expandafter \@firstoftwo
 \else \expandafter \@secondoftwo
 \fi
}%
\providecommand \@ifx [1]{%
 \ifx #1\expandafter \@firstoftwo
 \else \expandafter \@secondoftwo
 \fi
}%
\providecommand \natexlab [1]{#1}%
\providecommand \enquote  [1]{``#1''}%
\providecommand \bibnamefont  [1]{#1}%
\providecommand \bibfnamefont [1]{#1}%
\providecommand \citenamefont [1]{#1}%
\providecommand \href@noop [0]{\@secondoftwo}%
\providecommand \href [0]{\begingroup \@sanitize@url \@href}%
\providecommand \@href[1]{\@@startlink{#1}\@@href}%
\providecommand \@@href[1]{\endgroup#1\@@endlink}%
\providecommand \@sanitize@url [0]{\catcode `\\12\catcode `\$12\catcode
  `\&12\catcode `\#12\catcode `\^12\catcode `\_12\catcode `\%12\relax}%
\providecommand \@@startlink[1]{}%
\providecommand \@@endlink[0]{}%
\providecommand \url  [0]{\begingroup\@sanitize@url \@url }%
\providecommand \@url [1]{\endgroup\@href {#1}{\urlprefix }}%
\providecommand \urlprefix  [0]{URL }%
\providecommand \Eprint [0]{\href }%
\providecommand \doibase [0]{https://doi.org/}%
\providecommand \selectlanguage [0]{\@gobble}%
\providecommand \bibinfo  [0]{\@secondoftwo}%
\providecommand \bibfield  [0]{\@secondoftwo}%
\providecommand \translation [1]{[#1]}%
\providecommand \BibitemOpen [0]{}%
\providecommand \bibitemStop [0]{}%
\providecommand \bibitemNoStop [0]{.\EOS\space}%
\providecommand \EOS [0]{\spacefactor3000\relax}%
\providecommand \BibitemShut  [1]{\csname bibitem#1\endcsname}%
\let\auto@bib@innerbib\@empty
\bibitem [{\citenamefont {Berry}(1977)}]{Berry1977}%
  \BibitemOpen
  \bibfield  {author} {\bibinfo {author} {\bibfnamefont {M.~V.}\ \bibnamefont
  {Berry}},\ }\bibfield  {title} {\bibinfo {title} {Regular and irregular
  semiclassical wavefunctions},\ }\href
  {https://doi.org/10.1088/0305-4470/10/12/016} {\bibfield  {journal} {\bibinfo
   {journal} {J. Phys. A: Math. Gen.}\ }\textbf {\bibinfo {volume} {10}},\
  \bibinfo {pages} {2083} (\bibinfo {year} {1977})}\BibitemShut {NoStop}%
\bibitem [{\citenamefont {Deutsch}(1991)}]{Deutsch1991}%
  \BibitemOpen
  \bibfield  {author} {\bibinfo {author} {\bibfnamefont {J.~M.}\ \bibnamefont
  {Deutsch}},\ }\bibfield  {title} {\bibinfo {title} {Quantum statistical
  mechanics in a closed system},\ }\href
  {https://doi.org/10.1103/PhysRevA.43.2046} {\bibfield  {journal} {\bibinfo
  {journal} {Phys. Rev. A}\ }\textbf {\bibinfo {volume} {43}},\ \bibinfo
  {pages} {2046} (\bibinfo {year} {1991})}\BibitemShut {NoStop}%
\bibitem [{\citenamefont {Srednicki}(1994)}]{Srednicki1994}%
  \BibitemOpen
  \bibfield  {author} {\bibinfo {author} {\bibfnamefont {M.}~\bibnamefont
  {Srednicki}},\ }\bibfield  {title} {\bibinfo {title} {Chaos and quantum
  thermalization},\ }\href {https://doi.org/10.1103/PhysRevE.50.888} {\bibfield
   {journal} {\bibinfo  {journal} {Phys. Rev. E}\ }\textbf {\bibinfo {volume}
  {50}},\ \bibinfo {pages} {888} (\bibinfo {year} {1994})}\BibitemShut
  {NoStop}%
\bibitem [{\citenamefont {Rigol}\ \emph {et~al.}(2008)\citenamefont {Rigol},
  \citenamefont {Dunjko},\ and\ \citenamefont {Olshanii}}]{Rigol2008}%
  \BibitemOpen
  \bibfield  {author} {\bibinfo {author} {\bibfnamefont {M.}~\bibnamefont
  {Rigol}}, \bibinfo {author} {\bibfnamefont {V.}~\bibnamefont {Dunjko}},\ and\
  \bibinfo {author} {\bibfnamefont {M.}~\bibnamefont {Olshanii}},\ }\bibfield
  {title} {\bibinfo {title} {Thermalization and its mechanism for generic
  isolated quantum systems},\ }\href {https://doi.org/10.1038/nature06838}
  {\bibfield  {journal} {\bibinfo  {journal} {Nature}\ }\textbf {\bibinfo
  {volume} {452}},\ \bibinfo {pages} {854} (\bibinfo {year}
  {2008})}\BibitemShut {NoStop}%
\bibitem [{\citenamefont {D'Alessio}\ \emph {et~al.}(2016)\citenamefont
  {D'Alessio}, \citenamefont {Kafri}, \citenamefont {Polkovnikov},\ and\
  \citenamefont {Rigol}}]{DAlessio2016}%
  \BibitemOpen
  \bibfield  {author} {\bibinfo {author} {\bibfnamefont {L.}~\bibnamefont
  {D'Alessio}}, \bibinfo {author} {\bibfnamefont {Y.}~\bibnamefont {Kafri}},
  \bibinfo {author} {\bibfnamefont {A.}~\bibnamefont {Polkovnikov}},\ and\
  \bibinfo {author} {\bibfnamefont {M.}~\bibnamefont {Rigol}},\ }\bibfield
  {title} {\bibinfo {title} {From quantum chaos and eigenstate thermalization
  to statistical mechanics and thermodynamics},\ }\href
  {https://doi.org/10.1080/00018732.2016.1198134} {\bibfield  {journal}
  {\bibinfo  {journal} {Adv. Phys.}\ }\textbf {\bibinfo {volume} {65}},\
  \bibinfo {pages} {239} (\bibinfo {year} {2016})}\BibitemShut {NoStop}%
\bibitem [{\citenamefont {Gogolin}\ and\ \citenamefont
  {Eisert}(2016)}]{Gogolin2016}%
  \BibitemOpen
  \bibfield  {author} {\bibinfo {author} {\bibfnamefont {C.}~\bibnamefont
  {Gogolin}}\ and\ \bibinfo {author} {\bibfnamefont {J.}~\bibnamefont
  {Eisert}},\ }\bibfield  {title} {\bibinfo {title} {Equilibration,
  thermalisation, and the emergence of statistical mechanics in closed quantum
  systems},\ }\href {https://doi.org/10.1088/0034-4885/79/5/056001} {\bibfield
  {journal} {\bibinfo  {journal} {Rep. Prog. Phys.}\ }\textbf {\bibinfo
  {volume} {79}},\ \bibinfo {pages} {056001} (\bibinfo {year}
  {2016})}\BibitemShut {NoStop}%
\bibitem [{\citenamefont {Ueda}(2020)}]{Ueda2020}%
  \BibitemOpen
  \bibfield  {author} {\bibinfo {author} {\bibfnamefont {M.}~\bibnamefont
  {Ueda}},\ }\bibfield  {title} {\bibinfo {title} {Quantum equilibration,
  thermalization and prethermalization in ultracold atoms},\ }\href
  {https://doi.org/10.1038/s42254-020-0237-x} {\bibfield  {journal} {\bibinfo
  {journal} {Nat. Rev. Phys.}\ }\textbf {\bibinfo {volume} {2}},\ \bibinfo
  {pages} {669} (\bibinfo {year} {2020})}\BibitemShut {NoStop}%
\bibitem [{\citenamefont {Pilatowsky-Cameo}\ \emph {et~al.}(2023)\citenamefont
  {Pilatowsky-Cameo}, \citenamefont {Dag}, \citenamefont {Ho},\ and\
  \citenamefont {Choi}}]{PilatowskyCameo2023}%
  \BibitemOpen
  \bibfield  {author} {\bibinfo {author} {\bibfnamefont {S.}~\bibnamefont
  {Pilatowsky-Cameo}}, \bibinfo {author} {\bibfnamefont {C.~B.}\ \bibnamefont
  {Dag}}, \bibinfo {author} {\bibfnamefont {W.~W.}\ \bibnamefont {Ho}},\ and\
  \bibinfo {author} {\bibfnamefont {S.}~\bibnamefont {Choi}},\ }\bibfield
  {title} {\bibinfo {title} {Complete {Hilbert}-space ergodicity in quantum
  dynamics of generalized {Fibonacci} drives},\ }\href
  {https://doi.org/10.1103/PhysRevLett.131.250401} {\bibfield  {journal}
  {\bibinfo  {journal} {Phys. Rev. Lett.}\ }\textbf {\bibinfo {volume} {131}},\
  \bibinfo {pages} {250401} (\bibinfo {year} {2023})}\BibitemShut {NoStop}%
\bibitem [{\citenamefont {Pilatowsky-Cameo}\ \emph {et~al.}(2024)\citenamefont
  {Pilatowsky-Cameo}, \citenamefont {Marvian}, \citenamefont {Choi},\ and\
  \citenamefont {Ho}}]{PilatowskyCameo2024}%
  \BibitemOpen
  \bibfield  {author} {\bibinfo {author} {\bibfnamefont {S.}~\bibnamefont
  {Pilatowsky-Cameo}}, \bibinfo {author} {\bibfnamefont {I.}~\bibnamefont
  {Marvian}}, \bibinfo {author} {\bibfnamefont {S.}~\bibnamefont {Choi}},\ and\
  \bibinfo {author} {\bibfnamefont {W.~W.}\ \bibnamefont {Ho}},\ }\bibfield
  {title} {\bibinfo {title} {{Hilbert}-space ergodicity in driven quantum
  systems: Obstructions and designs},\ }\href
  {https://doi.org/10.1103/PhysRevX.14.041059} {\bibfield  {journal} {\bibinfo
  {journal} {Phys. Rev. X}\ }\textbf {\bibinfo {volume} {14}},\ \bibinfo
  {pages} {041059} (\bibinfo {year} {2024})}\BibitemShut {NoStop}%
\bibitem [{\citenamefont {Pilatowsky-Cameo}\ \emph {et~al.}(2025)\citenamefont
  {Pilatowsky-Cameo}, \citenamefont {Choi},\ and\ \citenamefont
  {Ho}}]{PilatowskyCameo2025}%
  \BibitemOpen
  \bibfield  {author} {\bibinfo {author} {\bibfnamefont {S.}~\bibnamefont
  {Pilatowsky-Cameo}}, \bibinfo {author} {\bibfnamefont {S.}~\bibnamefont
  {Choi}},\ and\ \bibinfo {author} {\bibfnamefont {W.~W.}\ \bibnamefont {Ho}},\
  }\bibfield  {title} {\bibinfo {title} {Critically slow {Hilbert}-space
  ergodicity in quantum morphic drives},\ }\href
  {https://doi.org/10.1103/dmfd-lgcq} {\bibfield  {journal} {\bibinfo
  {journal} {Phys. Rev. Lett.}\ }\textbf {\bibinfo {volume} {135}},\ \bibinfo
  {pages} {140402} (\bibinfo {year} {2025})}\BibitemShut {NoStop}%
\bibitem [{\citenamefont {Logari{\'c}}\ \emph {et~al.}(2025)\citenamefont
  {Logari{\'c}}, \citenamefont {Goold},\ and\ \citenamefont
  {Dooley}}]{Logaric2025}%
  \BibitemOpen
  \bibfield  {author} {\bibinfo {author} {\bibfnamefont {L.}~\bibnamefont
  {Logari{\'c}}}, \bibinfo {author} {\bibfnamefont {J.}~\bibnamefont {Goold}},\
  and\ \bibinfo {author} {\bibfnamefont {S.}~\bibnamefont {Dooley}},\
  }\bibfield  {title} {\bibinfo {title} {Hilbert subspace ergodicity},\ }\href
  {https://doi.org/10.1103/PhysRevB.111.144310} {\bibfield  {journal} {\bibinfo
   {journal} {Phys. Rev. B}\ }\textbf {\bibinfo {volume} {111}},\ \bibinfo
  {pages} {144310} (\bibinfo {year} {2025})}\BibitemShut {NoStop}%
\bibitem [{\citenamefont {Boltzmann}(1896)}]{Boltzmann1896}%
  \BibitemOpen
  \bibfield  {author} {\bibinfo {author} {\bibfnamefont {L.}~\bibnamefont
  {Boltzmann}},\ }\href@noop {} {\emph {\bibinfo {title} {Vorlesungen {\"u}ber
  {G}astheorie}}}\ (\bibinfo  {publisher} {Johann Ambrosius Barth},\ \bibinfo
  {address} {Leipzig},\ \bibinfo {year} {1896})\BibitemShut {NoStop}%
\bibitem [{\citenamefont {Birkhoff}(1931)}]{Birkhoff1931}%
  \BibitemOpen
  \bibfield  {author} {\bibinfo {author} {\bibfnamefont {G.~D.}\ \bibnamefont
  {Birkhoff}},\ }\bibfield  {title} {\bibinfo {title} {Proof of the ergodic
  theorem},\ }\href {https://doi.org/10.1073/pnas.17.2.656} {\bibfield
  {journal} {\bibinfo  {journal} {Proc. Natl. Acad. Sci. U.S.A.}\ }\textbf
  {\bibinfo {volume} {17}},\ \bibinfo {pages} {656} (\bibinfo {year}
  {1931})}\BibitemShut {NoStop}%
\bibitem [{\citenamefont {Lebowitz}\ and\ \citenamefont
  {Penrose}(1973)}]{LebowitzPenrose1973}%
  \BibitemOpen
  \bibfield  {author} {\bibinfo {author} {\bibfnamefont {J.~L.}\ \bibnamefont
  {Lebowitz}}\ and\ \bibinfo {author} {\bibfnamefont {O.}~\bibnamefont
  {Penrose}},\ }\bibfield  {title} {\bibinfo {title} {Modern ergodic theory},\
  }\href {https://doi.org/10.1063/1.3127948} {\bibfield  {journal} {\bibinfo
  {journal} {Phys. Today}\ }\textbf {\bibinfo {volume} {26}},\ \bibinfo {pages}
  {23} (\bibinfo {year} {1973})}\BibitemShut {NoStop}%
\bibitem [{\citenamefont {Cornfeld}\ \emph {et~al.}(1982)\citenamefont
  {Cornfeld}, \citenamefont {Fomin},\ and\ \citenamefont
  {Sinai}}]{CornfeldFominSinai1982}%
  \BibitemOpen
  \bibfield  {author} {\bibinfo {author} {\bibfnamefont {I.~P.}\ \bibnamefont
  {Cornfeld}}, \bibinfo {author} {\bibfnamefont {S.~V.}\ \bibnamefont
  {Fomin}},\ and\ \bibinfo {author} {\bibfnamefont {Y.~G.}\ \bibnamefont
  {Sinai}},\ }\href@noop {} {\emph {\bibinfo {title} {Ergodic Theory}}}\
  (\bibinfo  {publisher} {Springer},\ \bibinfo {address} {New York},\ \bibinfo
  {year} {1982})\BibitemShut {NoStop}%
\bibitem [{\citenamefont {Toda}\ \emph {et~al.}(1983)\citenamefont {Toda},
  \citenamefont {Kubo},\ and\ \citenamefont {Saito}}]{TodaKuboSaito}%
  \BibitemOpen
  \bibfield  {author} {\bibinfo {author} {\bibfnamefont {M.}~\bibnamefont
  {Toda}}, \bibinfo {author} {\bibfnamefont {R.}~\bibnamefont {Kubo}},\ and\
  \bibinfo {author} {\bibfnamefont {N.}~\bibnamefont {Saito}},\ }\href@noop {}
  {\emph {\bibinfo {title} {Statistical Physics {I}: Equilibrium Statistical
  Mechanics}}}\ (\bibinfo  {publisher} {Springer},\ \bibinfo {address} {Berlin,
  Heidelberg},\ \bibinfo {year} {1983})\BibitemShut {NoStop}%
\bibitem [{\citenamefont {Shaw}\ \emph {et~al.}(2025)\citenamefont {Shaw},
  \citenamefont {Mark}, \citenamefont {Choi}, \citenamefont {Finkelstein},
  \citenamefont {Scholl}, \citenamefont {Choi},\ and\ \citenamefont
  {Endres}}]{Shaw2025}%
  \BibitemOpen
  \bibfield  {author} {\bibinfo {author} {\bibfnamefont {A.~L.}\ \bibnamefont
  {Shaw}}, \bibinfo {author} {\bibfnamefont {D.~K.}\ \bibnamefont {Mark}},
  \bibinfo {author} {\bibfnamefont {J.}~\bibnamefont {Choi}}, \bibinfo {author}
  {\bibfnamefont {R.}~\bibnamefont {Finkelstein}}, \bibinfo {author}
  {\bibfnamefont {P.}~\bibnamefont {Scholl}}, \bibinfo {author} {\bibfnamefont
  {S.}~\bibnamefont {Choi}},\ and\ \bibinfo {author} {\bibfnamefont
  {M.}~\bibnamefont {Endres}},\ }\bibfield  {title} {\bibinfo {title}
  {Experimental signatures of {Hilbert}-space ergodicity: Universal bitstring
  distributions and applications in noise learning},\ }\href
  {https://doi.org/10.1103/h6xy-zpx4} {\bibfield  {journal} {\bibinfo
  {journal} {Phys. Rev. X}\ }\textbf {\bibinfo {volume} {15}},\ \bibinfo
  {pages} {031001} (\bibinfo {year} {2025})}\BibitemShut {NoStop}%
\bibitem [{\citenamefont {Liu}\ \emph {et~al.}(2026)\citenamefont {Liu},
  \citenamefont {Pan}, \citenamefont {Fu}, \citenamefont {Ho},\ and\
  \citenamefont {Rong}}]{Liu2026}%
  \BibitemOpen
  \bibfield  {author} {\bibinfo {author} {\bibfnamefont {W.}~\bibnamefont
  {Liu}}, \bibinfo {author} {\bibfnamefont {Z.-W.}\ \bibnamefont {Pan}},
  \bibinfo {author} {\bibfnamefont {Y.}~\bibnamefont {Fu}}, \bibinfo {author}
  {\bibfnamefont {W.~W.}\ \bibnamefont {Ho}},\ and\ \bibinfo {author}
  {\bibfnamefont {X.}~\bibnamefont {Rong}},\ }\bibfield  {title} {\bibinfo
  {title} {Observation of hierarchy of {Hilbert} space ergodicities in the
  quantum dynamics of a single spin},\ }\href
  {https://doi.org/10.1103/6msb-cxbc} {\bibfield  {journal} {\bibinfo
  {journal} {Phys. Rev. Lett.}\ }\textbf {\bibinfo {volume} {136}},\ \bibinfo
  {pages} {020401} (\bibinfo {year} {2026})}\BibitemShut {NoStop}%
\bibitem [{\citenamefont {Nielsen}\ and\ \citenamefont
  {Chuang}(2010)}]{NielsenChuang2010}%
  \BibitemOpen
  \bibfield  {author} {\bibinfo {author} {\bibfnamefont {M.~A.}\ \bibnamefont
  {Nielsen}}\ and\ \bibinfo {author} {\bibfnamefont {I.~L.}\ \bibnamefont
  {Chuang}},\ }\href@noop {} {\emph {\bibinfo {title} {Quantum Computation and
  Quantum Information}}}\ (\bibinfo  {publisher} {Cambridge University Press},\
  \bibinfo {address} {Cambridge},\ \bibinfo {year} {2010})\BibitemShut
  {NoStop}%
\bibitem [{\citenamefont {Watrous}(2018)}]{BookWatrous2018}%
  \BibitemOpen
  \bibfield  {author} {\bibinfo {author} {\bibfnamefont {J.}~\bibnamefont
  {Watrous}},\ }\href@noop {} {\emph {\bibinfo {title} {The Theory of Quantum
  Information}}}\ (\bibinfo  {publisher} {Cambridge University Press},\
  \bibinfo {address} {Cambridge, England},\ \bibinfo {year} {2018})\BibitemShut
  {NoStop}%
\bibitem [{\citenamefont {Martin}\ \emph {et~al.}(2017)\citenamefont {Martin},
  \citenamefont {Refael},\ and\ \citenamefont {Halperin}}]{Martin2017}%
  \BibitemOpen
  \bibfield  {author} {\bibinfo {author} {\bibfnamefont {I.}~\bibnamefont
  {Martin}}, \bibinfo {author} {\bibfnamefont {G.}~\bibnamefont {Refael}},\
  and\ \bibinfo {author} {\bibfnamefont {B.}~\bibnamefont {Halperin}},\
  }\bibfield  {title} {\bibinfo {title} {Topological frequency conversion in
  strongly driven quantum systems},\ }\href
  {https://doi.org/10.1103/PhysRevX.7.041008} {\bibfield  {journal} {\bibinfo
  {journal} {Phys. Rev. X}\ }\textbf {\bibinfo {volume} {7}},\ \bibinfo {pages}
  {041008} (\bibinfo {year} {2017})}\BibitemShut {NoStop}%
\bibitem [{\citenamefont {Dumitrescu}\ \emph {et~al.}(2018)\citenamefont
  {Dumitrescu}, \citenamefont {Vasseur},\ and\ \citenamefont
  {Potter}}]{Dumitrescu2018}%
  \BibitemOpen
  \bibfield  {author} {\bibinfo {author} {\bibfnamefont {P.~T.}\ \bibnamefont
  {Dumitrescu}}, \bibinfo {author} {\bibfnamefont {R.}~\bibnamefont
  {Vasseur}},\ and\ \bibinfo {author} {\bibfnamefont {A.~C.}\ \bibnamefont
  {Potter}},\ }\bibfield  {title} {\bibinfo {title} {Logarithmically slow
  relaxation in quasiperiodically driven random spin chains},\ }\href
  {https://doi.org/10.1103/PhysRevLett.120.070602} {\bibfield  {journal}
  {\bibinfo  {journal} {Phys. Rev. Lett.}\ }\textbf {\bibinfo {volume} {120}},\
  \bibinfo {pages} {070602} (\bibinfo {year} {2018})}\BibitemShut {NoStop}%
\bibitem [{\citenamefont {Else}\ \emph {et~al.}(2020)\citenamefont {Else},
  \citenamefont {Ho},\ and\ \citenamefont {Dumitrescu}}]{Else2020}%
  \BibitemOpen
  \bibfield  {author} {\bibinfo {author} {\bibfnamefont {D.~V.}\ \bibnamefont
  {Else}}, \bibinfo {author} {\bibfnamefont {W.~W.}\ \bibnamefont {Ho}},\ and\
  \bibinfo {author} {\bibfnamefont {P.~T.}\ \bibnamefont {Dumitrescu}},\
  }\bibfield  {title} {\bibinfo {title} {Long-lived interacting phases of
  matter protected by multiple time-translation symmetries in quasiperiodically
  driven systems},\ }\href {https://doi.org/10.1103/PhysRevX.10.021032}
  {\bibfield  {journal} {\bibinfo  {journal} {Phys. Rev. X}\ }\textbf {\bibinfo
  {volume} {10}},\ \bibinfo {pages} {021032} (\bibinfo {year}
  {2020})}\BibitemShut {NoStop}%
\bibitem [{\citenamefont {Long}\ \emph {et~al.}(2022)\citenamefont {Long},
  \citenamefont {Crowley},\ and\ \citenamefont {Chandran}}]{Long2022}%
  \BibitemOpen
  \bibfield  {author} {\bibinfo {author} {\bibfnamefont {D.~M.}\ \bibnamefont
  {Long}}, \bibinfo {author} {\bibfnamefont {P.~J.~D.}\ \bibnamefont
  {Crowley}},\ and\ \bibinfo {author} {\bibfnamefont {A.}~\bibnamefont
  {Chandran}},\ }\bibfield  {title} {\bibinfo {title} {Many-body localization
  with quasiperiodic driving},\ }\href
  {https://doi.org/10.1103/PhysRevB.105.144204} {\bibfield  {journal} {\bibinfo
   {journal} {Phys. Rev. B}\ }\textbf {\bibinfo {volume} {105}},\ \bibinfo
  {pages} {144204} (\bibinfo {year} {2022})}\BibitemShut {NoStop}%
\bibitem [{\citenamefont {He}\ \emph {et~al.}(2025)\citenamefont {He},
  \citenamefont {Ye}, \citenamefont {Gong}, \citenamefont {Yao}, \citenamefont
  {Liu}, \citenamefont {Murch}, \citenamefont {Yao},\ and\ \citenamefont
  {Zu}}]{He2025}%
  \BibitemOpen
  \bibfield  {author} {\bibinfo {author} {\bibfnamefont {G.}~\bibnamefont
  {He}}, \bibinfo {author} {\bibfnamefont {B.}~\bibnamefont {Ye}}, \bibinfo
  {author} {\bibfnamefont {R.}~\bibnamefont {Gong}}, \bibinfo {author}
  {\bibfnamefont {C.}~\bibnamefont {Yao}}, \bibinfo {author} {\bibfnamefont
  {Z.}~\bibnamefont {Liu}}, \bibinfo {author} {\bibfnamefont {K.~W.}\
  \bibnamefont {Murch}}, \bibinfo {author} {\bibfnamefont {N.~Y.}\ \bibnamefont
  {Yao}},\ and\ \bibinfo {author} {\bibfnamefont {C.}~\bibnamefont {Zu}},\
  }\bibfield  {title} {\bibinfo {title} {Experimental realization of discrete
  time quasicrystals},\ }\href {https://doi.org/10.1103/PhysRevX.15.011055}
  {\bibfield  {journal} {\bibinfo  {journal} {Phys. Rev. X}\ }\textbf {\bibinfo
  {volume} {15}},\ \bibinfo {pages} {011055} (\bibinfo {year}
  {2025})}\BibitemShut {NoStop}%
\bibitem [{\citenamefont {Weyl}(1916)}]{Weyl1916}%
  \BibitemOpen
  \bibfield  {author} {\bibinfo {author} {\bibfnamefont {H.}~\bibnamefont
  {Weyl}},\ }\bibfield  {title} {\bibinfo {title} {{\"U}ber die
  {G}leichverteilung von {Z}ahlen mod.\ {E}ins},\ }\href
  {https://doi.org/10.1007/BF01475864} {\bibfield  {journal} {\bibinfo
  {journal} {Math. Ann.}\ }\textbf {\bibinfo {volume} {77}},\ \bibinfo {pages}
  {313} (\bibinfo {year} {1916})}\BibitemShut {NoStop}%
\bibitem [{SM()}]{SM}%
  \BibitemOpen
  \href@noop {} {}\bibinfo {note} {See Supplemental Material at [URL] for the
  proof of the $k$-HSE conditions of the $m$-tone drives and their invariance
  under permutations of the inter-axis angles, additional drive examples, the
  closed form of the $(k{+}1)$th-order deviation, and details of the diamond
  sample and experimental methods, which includes
  Ref.~\cite{Harrow2013}}\BibitemShut {NoStop}%
\bibitem [{\citenamefont {Bogaert}(2014)}]{Bogaert2014}%
  \BibitemOpen
  \bibfield  {author} {\bibinfo {author} {\bibfnamefont {I.}~\bibnamefont
  {Bogaert}},\ }\bibfield  {title} {\bibinfo {title} {Iteration-free
  computation of {Gauss}--{Legendre} quadrature nodes and weights},\ }\href
  {https://doi.org/10.1137/140954969} {\bibfield  {journal} {\bibinfo
  {journal} {SIAM J. Sci. Comput.}\ }\textbf {\bibinfo {volume} {36}},\
  \bibinfo {pages} {A1008} (\bibinfo {year} {2014})}\BibitemShut {NoStop}%
\bibitem [{\citenamefont {Gali}\ \emph {et~al.}(2008)\citenamefont {Gali},
  \citenamefont {Fyta},\ and\ \citenamefont {Kaxiras}}]{Gali2008}%
  \BibitemOpen
  \bibfield  {author} {\bibinfo {author} {\bibfnamefont {A.}~\bibnamefont
  {Gali}}, \bibinfo {author} {\bibfnamefont {M.}~\bibnamefont {Fyta}},\ and\
  \bibinfo {author} {\bibfnamefont {E.}~\bibnamefont {Kaxiras}},\ }\bibfield
  {title} {\bibinfo {title} {Ab initio supercell calculations on
  nitrogen-vacancy center in diamond: Electronic structure and hyperfine
  tensors},\ }\href {https://doi.org/10.1103/PhysRevB.77.155206} {\bibfield
  {journal} {\bibinfo  {journal} {Phys. Rev. B}\ }\textbf {\bibinfo {volume}
  {77}},\ \bibinfo {pages} {155206} (\bibinfo {year} {2008})}\BibitemShut
  {NoStop}%
\bibitem [{\citenamefont {Maze}\ \emph {et~al.}(2011)\citenamefont {Maze},
  \citenamefont {Gali}, \citenamefont {Togan}, \citenamefont {Chu},
  \citenamefont {Trifonov}, \citenamefont {Kaxiras},\ and\ \citenamefont
  {Lukin}}]{Maze2011}%
  \BibitemOpen
  \bibfield  {author} {\bibinfo {author} {\bibfnamefont {J.~R.}\ \bibnamefont
  {Maze}}, \bibinfo {author} {\bibfnamefont {A.}~\bibnamefont {Gali}}, \bibinfo
  {author} {\bibfnamefont {E.}~\bibnamefont {Togan}}, \bibinfo {author}
  {\bibfnamefont {Y.}~\bibnamefont {Chu}}, \bibinfo {author} {\bibfnamefont
  {A.}~\bibnamefont {Trifonov}}, \bibinfo {author} {\bibfnamefont
  {E.}~\bibnamefont {Kaxiras}},\ and\ \bibinfo {author} {\bibfnamefont {M.~D.}\
  \bibnamefont {Lukin}},\ }\bibfield  {title} {\bibinfo {title} {Properties of
  nitrogen-vacancy centers in diamond: The group theoretic approach},\ }\href
  {https://doi.org/10.1088/1367-2630/13/2/025025} {\bibfield  {journal}
  {\bibinfo  {journal} {New J. Phys.}\ }\textbf {\bibinfo {volume} {13}},\
  \bibinfo {pages} {025025} (\bibinfo {year} {2011})}\BibitemShut {NoStop}%
\bibitem [{\citenamefont {Doherty}\ \emph {et~al.}(2013)\citenamefont
  {Doherty}, \citenamefont {Manson}, \citenamefont {Delaney}, \citenamefont
  {Jelezko}, \citenamefont {Wrachtrup},\ and\ \citenamefont
  {Hollenberg}}]{Doherty2013}%
  \BibitemOpen
  \bibfield  {author} {\bibinfo {author} {\bibfnamefont {M.~W.}\ \bibnamefont
  {Doherty}}, \bibinfo {author} {\bibfnamefont {N.~B.}\ \bibnamefont {Manson}},
  \bibinfo {author} {\bibfnamefont {P.}~\bibnamefont {Delaney}}, \bibinfo
  {author} {\bibfnamefont {F.}~\bibnamefont {Jelezko}}, \bibinfo {author}
  {\bibfnamefont {J.}~\bibnamefont {Wrachtrup}},\ and\ \bibinfo {author}
  {\bibfnamefont {L.~C.~L.}\ \bibnamefont {Hollenberg}},\ }\bibfield  {title}
  {\bibinfo {title} {The nitrogen-vacancy colour centre in diamond},\ }\href
  {https://doi.org/10.1016/j.physrep.2013.02.001} {\bibfield  {journal}
  {\bibinfo  {journal} {Phys. Rep.}\ }\textbf {\bibinfo {volume} {528}},\
  \bibinfo {pages} {1} (\bibinfo {year} {2013})}\BibitemShut {NoStop}%
\bibitem [{\citenamefont {Jacques}\ \emph {et~al.}(2009)\citenamefont
  {Jacques}, \citenamefont {Neumann}, \citenamefont {Beck}, \citenamefont
  {Markham}, \citenamefont {Twitchen}, \citenamefont {Meijer}, \citenamefont
  {Kaiser}, \citenamefont {Balasubramanian}, \citenamefont {Jelezko},\ and\
  \citenamefont {Wrachtrup}}]{Jacques2009}%
  \BibitemOpen
  \bibfield  {author} {\bibinfo {author} {\bibfnamefont {V.}~\bibnamefont
  {Jacques}}, \bibinfo {author} {\bibfnamefont {P.}~\bibnamefont {Neumann}},
  \bibinfo {author} {\bibfnamefont {J.}~\bibnamefont {Beck}}, \bibinfo {author}
  {\bibfnamefont {M.}~\bibnamefont {Markham}}, \bibinfo {author} {\bibfnamefont
  {D.}~\bibnamefont {Twitchen}}, \bibinfo {author} {\bibfnamefont
  {J.}~\bibnamefont {Meijer}}, \bibinfo {author} {\bibfnamefont
  {F.}~\bibnamefont {Kaiser}}, \bibinfo {author} {\bibfnamefont
  {G.}~\bibnamefont {Balasubramanian}}, \bibinfo {author} {\bibfnamefont
  {F.}~\bibnamefont {Jelezko}},\ and\ \bibinfo {author} {\bibfnamefont
  {J.}~\bibnamefont {Wrachtrup}},\ }\bibfield  {title} {\bibinfo {title}
  {Dynamic polarization of single nuclear spins by optical pumping of
  nitrogen-vacancy color centers in diamond at room temperature},\ }\href
  {https://doi.org/10.1103/PhysRevLett.102.057403} {\bibfield  {journal}
  {\bibinfo  {journal} {Phys. Rev. Lett.}\ }\textbf {\bibinfo {volume} {102}},\
  \bibinfo {pages} {057403} (\bibinfo {year} {2009})}\BibitemShut {NoStop}%
\bibitem [{\citenamefont {Busaite}\ \emph {et~al.}(2020)\citenamefont
  {Busaite}, \citenamefont {Lazda}, \citenamefont {Berzins}, \citenamefont
  {Auzinsh}, \citenamefont {Ferber},\ and\ \citenamefont
  {Gahbauer}}]{Busaite2020}%
  \BibitemOpen
  \bibfield  {author} {\bibinfo {author} {\bibfnamefont {L.}~\bibnamefont
  {Busaite}}, \bibinfo {author} {\bibfnamefont {R.}~\bibnamefont {Lazda}},
  \bibinfo {author} {\bibfnamefont {A.}~\bibnamefont {Berzins}}, \bibinfo
  {author} {\bibfnamefont {M.}~\bibnamefont {Auzinsh}}, \bibinfo {author}
  {\bibfnamefont {R.}~\bibnamefont {Ferber}},\ and\ \bibinfo {author}
  {\bibfnamefont {F.}~\bibnamefont {Gahbauer}},\ }\bibfield  {title} {\bibinfo
  {title} {Dynamic $^{14}${N} nuclear spin polarization in nitrogen-vacancy
  centers in diamond},\ }\href {https://doi.org/10.1103/PhysRevB.102.224101}
  {\bibfield  {journal} {\bibinfo  {journal} {Phys. Rev. B}\ }\textbf {\bibinfo
  {volume} {102}},\ \bibinfo {pages} {224101} (\bibinfo {year}
  {2020})}\BibitemShut {NoStop}%
\bibitem [{\citenamefont {de~Lange}\ \emph {et~al.}(2010)\citenamefont
  {de~Lange}, \citenamefont {Wang}, \citenamefont {Rist\`{e}}, \citenamefont
  {Dobrovitski},\ and\ \citenamefont {Hanson}}]{deLange2010}%
  \BibitemOpen
  \bibfield  {author} {\bibinfo {author} {\bibfnamefont {G.}~\bibnamefont
  {de~Lange}}, \bibinfo {author} {\bibfnamefont {Z.~H.}\ \bibnamefont {Wang}},
  \bibinfo {author} {\bibfnamefont {D.}~\bibnamefont {Rist\`{e}}}, \bibinfo
  {author} {\bibfnamefont {V.~V.}\ \bibnamefont {Dobrovitski}},\ and\ \bibinfo
  {author} {\bibfnamefont {R.}~\bibnamefont {Hanson}},\ }\bibfield  {title}
  {\bibinfo {title} {Universal dynamical decoupling of a single solid-state
  spin from a spin bath},\ }\href {https://doi.org/10.1126/science.1192739}
  {\bibfield  {journal} {\bibinfo  {journal} {Science}\ }\textbf {\bibinfo
  {volume} {330}},\ \bibinfo {pages} {60} (\bibinfo {year} {2010})}\BibitemShut
  {NoStop}%
\bibitem [{\citenamefont {Choi}\ \emph {et~al.}(2017)\citenamefont {Choi},
  \citenamefont {Choi}, \citenamefont {Landig}, \citenamefont {Kucsko},
  \citenamefont {Zhou}, \citenamefont {Isoya}, \citenamefont {Jelezko},
  \citenamefont {Onoda}, \citenamefont {Sumiya}, \citenamefont {Khemani},
  \citenamefont {von Keyserlingk}, \citenamefont {Yao}, \citenamefont
  {Demler},\ and\ \citenamefont {Lukin}}]{Choi2017}%
  \BibitemOpen
  \bibfield  {author} {\bibinfo {author} {\bibfnamefont {S.}~\bibnamefont
  {Choi}}, \bibinfo {author} {\bibfnamefont {J.}~\bibnamefont {Choi}}, \bibinfo
  {author} {\bibfnamefont {R.}~\bibnamefont {Landig}}, \bibinfo {author}
  {\bibfnamefont {G.}~\bibnamefont {Kucsko}}, \bibinfo {author} {\bibfnamefont
  {H.}~\bibnamefont {Zhou}}, \bibinfo {author} {\bibfnamefont {J.}~\bibnamefont
  {Isoya}}, \bibinfo {author} {\bibfnamefont {F.}~\bibnamefont {Jelezko}},
  \bibinfo {author} {\bibfnamefont {S.}~\bibnamefont {Onoda}}, \bibinfo
  {author} {\bibfnamefont {H.}~\bibnamefont {Sumiya}}, \bibinfo {author}
  {\bibfnamefont {V.}~\bibnamefont {Khemani}}, \bibinfo {author} {\bibfnamefont
  {C.}~\bibnamefont {von Keyserlingk}}, \bibinfo {author} {\bibfnamefont
  {N.~Y.}\ \bibnamefont {Yao}}, \bibinfo {author} {\bibfnamefont
  {E.}~\bibnamefont {Demler}},\ and\ \bibinfo {author} {\bibfnamefont {M.~D.}\
  \bibnamefont {Lukin}},\ }\bibfield  {title} {\bibinfo {title} {Observation of
  discrete time-crystalline order in a disordered dipolar many-body system},\
  }\href {https://doi.org/10.1038/nature21426} {\bibfield  {journal} {\bibinfo
  {journal} {Nature}\ }\textbf {\bibinfo {volume} {543}},\ \bibinfo {pages}
  {221} (\bibinfo {year} {2017})}\BibitemShut {NoStop}%
\bibitem [{\citenamefont {Mark}\ \emph {et~al.}(2024)\citenamefont {Mark},
  \citenamefont {Surace}, \citenamefont {Elben}, \citenamefont {Shaw},
  \citenamefont {Choi}, \citenamefont {Refael}, \citenamefont {Endres},\ and\
  \citenamefont {Choi}}]{Mark2024}%
  \BibitemOpen
  \bibfield  {author} {\bibinfo {author} {\bibfnamefont {D.~K.}\ \bibnamefont
  {Mark}}, \bibinfo {author} {\bibfnamefont {F.}~\bibnamefont {Surace}},
  \bibinfo {author} {\bibfnamefont {A.}~\bibnamefont {Elben}}, \bibinfo
  {author} {\bibfnamefont {A.~L.}\ \bibnamefont {Shaw}}, \bibinfo {author}
  {\bibfnamefont {J.}~\bibnamefont {Choi}}, \bibinfo {author} {\bibfnamefont
  {G.}~\bibnamefont {Refael}}, \bibinfo {author} {\bibfnamefont
  {M.}~\bibnamefont {Endres}},\ and\ \bibinfo {author} {\bibfnamefont
  {S.}~\bibnamefont {Choi}},\ }\bibfield  {title} {\bibinfo {title} {Maximum
  entropy principle in deep thermalization and in {Hilbert}-space ergodicity},\
  }\href {https://doi.org/10.1103/PhysRevX.14.041051} {\bibfield  {journal}
  {\bibinfo  {journal} {Phys. Rev. X}\ }\textbf {\bibinfo {volume} {14}},\
  \bibinfo {pages} {041051} (\bibinfo {year} {2024})}\BibitemShut {NoStop}%
\bibitem [{\citenamefont {Mok}\ \emph {et~al.}()\citenamefont {Mok},
  \citenamefont {Haug}, \citenamefont {Ho},\ and\ \citenamefont
  {Preskill}}]{MokScrooge2026}%
  \BibitemOpen
  \bibfield  {author} {\bibinfo {author} {\bibfnamefont {W.-K.}\ \bibnamefont
  {Mok}}, \bibinfo {author} {\bibfnamefont {T.}~\bibnamefont {Haug}}, \bibinfo
  {author} {\bibfnamefont {W.~W.}\ \bibnamefont {Ho}},\ and\ \bibinfo {author}
  {\bibfnamefont {J.}~\bibnamefont {Preskill}},\ }\bibfield  {title} {\bibinfo
  {title} {Nature is stingy: Universality of {Scrooge} ensembles in quantum
  many-body systems},\ }\Eprint {https://arxiv.org/abs/2601.00266}
  {arXiv:2601.00266} \BibitemShut {NoStop}%
\bibitem [{\citenamefont {Ho}\ and\ \citenamefont {Choi}(2022)}]{Ho2022}%
  \BibitemOpen
  \bibfield  {author} {\bibinfo {author} {\bibfnamefont {W.~W.}\ \bibnamefont
  {Ho}}\ and\ \bibinfo {author} {\bibfnamefont {S.}~\bibnamefont {Choi}},\
  }\bibfield  {title} {\bibinfo {title} {Exact emergent quantum state designs
  from quantum chaotic dynamics},\ }\href
  {https://doi.org/10.1103/PhysRevLett.128.060601} {\bibfield  {journal}
  {\bibinfo  {journal} {Phys. Rev. Lett.}\ }\textbf {\bibinfo {volume} {128}},\
  \bibinfo {pages} {060601} (\bibinfo {year} {2022})}\BibitemShut {NoStop}%
\bibitem [{\citenamefont {Ippoliti}\ and\ \citenamefont
  {Ho}(2022)}]{Ippoliti2022}%
  \BibitemOpen
  \bibfield  {author} {\bibinfo {author} {\bibfnamefont {M.}~\bibnamefont
  {Ippoliti}}\ and\ \bibinfo {author} {\bibfnamefont {W.~W.}\ \bibnamefont
  {Ho}},\ }\bibfield  {title} {\bibinfo {title} {Solvable model of deep
  thermalization with distinct design times},\ }\href
  {https://doi.org/10.22331/q-2022-12-29-886} {\bibfield  {journal} {\bibinfo
  {journal} {Quantum}\ }\textbf {\bibinfo {volume} {6}},\ \bibinfo {pages}
  {886} (\bibinfo {year} {2022})}\BibitemShut {NoStop}%
\bibitem [{\citenamefont {Choi}\ \emph {et~al.}(2023)\citenamefont {Choi},
  \citenamefont {Shaw}, \citenamefont {Madjarov}, \citenamefont {Xie},
  \citenamefont {Finkelstein}, \citenamefont {Covey}, \citenamefont {Cotler},
  \citenamefont {Mark}, \citenamefont {Huang}, \citenamefont {Kale},
  \citenamefont {Pichler}, \citenamefont {Brand{\~a}o}, \citenamefont {Choi},\
  and\ \citenamefont {Endres}}]{Choi2023}%
  \BibitemOpen
  \bibfield  {author} {\bibinfo {author} {\bibfnamefont {J.}~\bibnamefont
  {Choi}}, \bibinfo {author} {\bibfnamefont {A.~L.}\ \bibnamefont {Shaw}},
  \bibinfo {author} {\bibfnamefont {I.~S.}\ \bibnamefont {Madjarov}}, \bibinfo
  {author} {\bibfnamefont {X.}~\bibnamefont {Xie}}, \bibinfo {author}
  {\bibfnamefont {R.}~\bibnamefont {Finkelstein}}, \bibinfo {author}
  {\bibfnamefont {J.~P.}\ \bibnamefont {Covey}}, \bibinfo {author}
  {\bibfnamefont {J.~S.}\ \bibnamefont {Cotler}}, \bibinfo {author}
  {\bibfnamefont {D.~K.}\ \bibnamefont {Mark}}, \bibinfo {author}
  {\bibfnamefont {H.-Y.}\ \bibnamefont {Huang}}, \bibinfo {author}
  {\bibfnamefont {A.}~\bibnamefont {Kale}}, \bibinfo {author} {\bibfnamefont
  {H.}~\bibnamefont {Pichler}}, \bibinfo {author} {\bibfnamefont {F.~G. S.~L.}\
  \bibnamefont {Brand{\~a}o}}, \bibinfo {author} {\bibfnamefont
  {S.}~\bibnamefont {Choi}},\ and\ \bibinfo {author} {\bibfnamefont
  {M.}~\bibnamefont {Endres}},\ }\bibfield  {title} {\bibinfo {title}
  {Preparing random states and benchmarking with many-body quantum chaos},\
  }\href {https://doi.org/10.1038/s41586-022-05442-1} {\bibfield  {journal}
  {\bibinfo  {journal} {Nature}\ }\textbf {\bibinfo {volume} {613}},\ \bibinfo
  {pages} {468} (\bibinfo {year} {2023})}\BibitemShut {NoStop}%
\bibitem [{\citenamefont {Cotler}\ \emph {et~al.}(2023)\citenamefont {Cotler},
  \citenamefont {Mark}, \citenamefont {Huang}, \citenamefont {Hern{\'a}ndez},
  \citenamefont {Choi}, \citenamefont {Shaw}, \citenamefont {Endres},\ and\
  \citenamefont {Choi}}]{Cotler2023}%
  \BibitemOpen
  \bibfield  {author} {\bibinfo {author} {\bibfnamefont {J.~S.}\ \bibnamefont
  {Cotler}}, \bibinfo {author} {\bibfnamefont {D.~K.}\ \bibnamefont {Mark}},
  \bibinfo {author} {\bibfnamefont {H.-Y.}\ \bibnamefont {Huang}}, \bibinfo
  {author} {\bibfnamefont {F.}~\bibnamefont {Hern{\'a}ndez}}, \bibinfo {author}
  {\bibfnamefont {J.}~\bibnamefont {Choi}}, \bibinfo {author} {\bibfnamefont
  {A.~L.}\ \bibnamefont {Shaw}}, \bibinfo {author} {\bibfnamefont
  {M.}~\bibnamefont {Endres}},\ and\ \bibinfo {author} {\bibfnamefont
  {S.}~\bibnamefont {Choi}},\ }\bibfield  {title} {\bibinfo {title} {Emergent
  quantum state designs from individual many-body wave functions},\ }\href
  {https://doi.org/10.1103/PRXQuantum.4.010311} {\bibfield  {journal} {\bibinfo
   {journal} {PRX Quantum}\ }\textbf {\bibinfo {volume} {4}},\ \bibinfo {pages}
  {010311} (\bibinfo {year} {2023})}\BibitemShut {NoStop}%
\bibitem [{\citenamefont {Liu}\ \emph {et~al.}(2024)\citenamefont {Liu},
  \citenamefont {Huang},\ and\ \citenamefont {Ho}}]{Liu2024}%
  \BibitemOpen
  \bibfield  {author} {\bibinfo {author} {\bibfnamefont {C.}~\bibnamefont
  {Liu}}, \bibinfo {author} {\bibfnamefont {Q.~C.}\ \bibnamefont {Huang}},\
  and\ \bibinfo {author} {\bibfnamefont {W.~W.}\ \bibnamefont {Ho}},\
  }\bibfield  {title} {\bibinfo {title} {Deep thermalization in {Gaussian}
  continuous-variable quantum systems},\ }\href
  {https://doi.org/10.1103/PhysRevLett.133.260401} {\bibfield  {journal}
  {\bibinfo  {journal} {Phys. Rev. Lett.}\ }\textbf {\bibinfo {volume} {133}},\
  \bibinfo {pages} {260401} (\bibinfo {year} {2024})}\BibitemShut {NoStop}%
\bibitem [{\citenamefont {Chang}\ \emph {et~al.}(2025)\citenamefont {Chang},
  \citenamefont {Shrotriya}, \citenamefont {Ho},\ and\ \citenamefont
  {Ippoliti}}]{Chang2025}%
  \BibitemOpen
  \bibfield  {author} {\bibinfo {author} {\bibfnamefont {R.-A.}\ \bibnamefont
  {Chang}}, \bibinfo {author} {\bibfnamefont {H.}~\bibnamefont {Shrotriya}},
  \bibinfo {author} {\bibfnamefont {W.~W.}\ \bibnamefont {Ho}},\ and\ \bibinfo
  {author} {\bibfnamefont {M.}~\bibnamefont {Ippoliti}},\ }\bibfield  {title}
  {\bibinfo {title} {Deep thermalization under charge-conserving quantum
  dynamics},\ }\href {https://doi.org/10.1103/PRXQuantum.6.020343} {\bibfield
  {journal} {\bibinfo  {journal} {PRX Quantum}\ }\textbf {\bibinfo {volume}
  {6}},\ \bibinfo {pages} {020343} (\bibinfo {year} {2025})}\BibitemShut
  {NoStop}%
\bibitem [{\citenamefont {Chakraborty}\ \emph {et~al.}(2025)\citenamefont
  {Chakraborty}, \citenamefont {Choi}, \citenamefont {Ghosh},\ and\
  \citenamefont {Giurgic{\u{a}}-Tiron}}]{Chakraborty2025}%
  \BibitemOpen
  \bibfield  {author} {\bibinfo {author} {\bibfnamefont {S.}~\bibnamefont
  {Chakraborty}}, \bibinfo {author} {\bibfnamefont {S.}~\bibnamefont {Choi}},
  \bibinfo {author} {\bibfnamefont {S.}~\bibnamefont {Ghosh}},\ and\ \bibinfo
  {author} {\bibfnamefont {T.}~\bibnamefont {Giurgic{\u{a}}-Tiron}},\
  }\bibfield  {title} {\bibinfo {title} {Fast computational deep
  thermalization},\ }\href {https://doi.org/10.1103/g3qz-b2lv} {\bibfield
  {journal} {\bibinfo  {journal} {Phys. Rev. Lett.}\ }\textbf {\bibinfo
  {volume} {135}},\ \bibinfo {pages} {210603} (\bibinfo {year}
  {2025})}\BibitemShut {NoStop}%
\bibitem [{\citenamefont {Harrow}()}]{Harrow2013}%
  \BibitemOpen
  \bibfield  {author} {\bibinfo {author} {\bibfnamefont {A.~W.}\ \bibnamefont
  {Harrow}},\ }\bibfield  {title} {\bibinfo {title} {The church of the
  symmetric subspace},\ }\Eprint {https://arxiv.org/abs/1308.6595}
  {arXiv:1308.6595} \BibitemShut {NoStop}%
\end{thebibliography}%


\begin{thebibliography}{6}%
\makeatletter
\providecommand \@ifxundefined [1]{%
 \@ifx{#1\undefined}
}%
\providecommand \@ifnum [1]{%
 \ifnum #1\expandafter \@firstoftwo
 \else \expandafter \@secondoftwo
 \fi
}%
\providecommand \@ifx [1]{%
 \ifx #1\expandafter \@firstoftwo
 \else \expandafter \@secondoftwo
 \fi
}%
\providecommand \natexlab [1]{#1}%
\providecommand \enquote  [1]{``#1''}%
\providecommand \bibnamefont  [1]{#1}%
\providecommand \bibfnamefont [1]{#1}%
\providecommand \citenamefont [1]{#1}%
\providecommand \href@noop [0]{\@secondoftwo}%
\providecommand \href [0]{\begingroup \@sanitize@url \@href}%
\providecommand \@href[1]{\@@startlink{#1}\@@href}%
\providecommand \@@href[1]{\endgroup#1\@@endlink}%
\providecommand \@sanitize@url [0]{\catcode `\\12\catcode `\$12\catcode
  `\&12\catcode `\#12\catcode `\^12\catcode `\_12\catcode `\%12\relax}%
\providecommand \@@startlink[1]{}%
\providecommand \@@endlink[0]{}%
\providecommand \url  [0]{\begingroup\@sanitize@url \@url }%
\providecommand \@url [1]{\endgroup\@href {#1}{\urlprefix }}%
\providecommand \urlprefix  [0]{URL }%
\providecommand \Eprint [0]{\href }%
\providecommand \doibase [0]{https://doi.org/}%
\providecommand \selectlanguage [0]{\@gobble}%
\providecommand \bibinfo  [0]{\@secondoftwo}%
\providecommand \bibfield  [0]{\@secondoftwo}%
\providecommand \translation [1]{[#1]}%
\providecommand \BibitemOpen [0]{}%
\providecommand \bibitemStop [0]{}%
\providecommand \bibitemNoStop [0]{.\EOS\space}%
\providecommand \EOS [0]{\spacefactor3000\relax}%
\providecommand \BibitemShut  [1]{\csname bibitem#1\endcsname}%
\let\auto@bib@innerbib\@empty
\bibitem [{\citenamefont {Pilatowsky-Cameo}\ \emph {et~al.}(2024)\citenamefont
  {Pilatowsky-Cameo}, \citenamefont {Marvian}, \citenamefont {Choi},\ and\
  \citenamefont {Ho}}]{SM-PilatowskyCameo2024}%
  \BibitemOpen
  \bibfield  {author} {\bibinfo {author} {\bibfnamefont {S.}~\bibnamefont
  {Pilatowsky-Cameo}}, \bibinfo {author} {\bibfnamefont {I.}~\bibnamefont
  {Marvian}}, \bibinfo {author} {\bibfnamefont {S.}~\bibnamefont {Choi}},\ and\
  \bibinfo {author} {\bibfnamefont {W.~W.}\ \bibnamefont {Ho}},\ }\bibfield
  {title} {\bibinfo {title} {{Hilbert}-space ergodicity in driven quantum
  systems: Obstructions and designs},\ }\href
  {https://doi.org/10.1103/PhysRevX.14.041059} {\bibfield  {journal} {\bibinfo
  {journal} {Phys. Rev. X}\ }\textbf {\bibinfo {volume} {14}},\ \bibinfo
  {pages} {041059} (\bibinfo {year} {2024})}\BibitemShut {NoStop}%
\bibitem [{\citenamefont {Harrow}()}]{SM-Harrow2013}%
  \BibitemOpen
  \bibfield  {author} {\bibinfo {author} {\bibfnamefont {A.~W.}\ \bibnamefont
  {Harrow}},\ }\bibfield  {title} {\bibinfo {title} {The church of the
  symmetric subspace},\ }\Eprint {https://arxiv.org/abs/1308.6595}
  {arXiv:1308.6595} \BibitemShut {NoStop}%
\bibitem [{\citenamefont {Weyl}(1916)}]{SM-Weyl1916}%
  \BibitemOpen
  \bibfield  {author} {\bibinfo {author} {\bibfnamefont {H.}~\bibnamefont
  {Weyl}},\ }\bibfield  {title} {\bibinfo {title} {{\"U}ber die
  {G}leichverteilung von {Z}ahlen mod.\ {E}ins},\ }\href
  {https://doi.org/10.1007/BF01475864} {\bibfield  {journal} {\bibinfo
  {journal} {Math. Ann.}\ }\textbf {\bibinfo {volume} {77}},\ \bibinfo {pages}
  {313} (\bibinfo {year} {1916})}\BibitemShut {NoStop}%
\bibitem [{\citenamefont {Bogaert}(2014)}]{SM-Bogaert2014}%
  \BibitemOpen
  \bibfield  {author} {\bibinfo {author} {\bibfnamefont {I.}~\bibnamefont
  {Bogaert}},\ }\bibfield  {title} {\bibinfo {title} {Iteration-free
  computation of {Gauss}--{Legendre} quadrature nodes and weights},\ }\href
  {https://doi.org/10.1137/140954969} {\bibfield  {journal} {\bibinfo
  {journal} {SIAM J. Sci. Comput.}\ }\textbf {\bibinfo {volume} {36}},\
  \bibinfo {pages} {A1008} (\bibinfo {year} {2014})}\BibitemShut {NoStop}%
\bibitem [{\citenamefont {Liu}\ \emph {et~al.}(2026)\citenamefont {Liu},
  \citenamefont {Pan}, \citenamefont {Fu}, \citenamefont {Ho},\ and\
  \citenamefont {Rong}}]{SM-Liu2026}%
  \BibitemOpen
  \bibfield  {author} {\bibinfo {author} {\bibfnamefont {W.}~\bibnamefont
  {Liu}}, \bibinfo {author} {\bibfnamefont {Z.-W.}\ \bibnamefont {Pan}},
  \bibinfo {author} {\bibfnamefont {Y.}~\bibnamefont {Fu}}, \bibinfo {author}
  {\bibfnamefont {W.~W.}\ \bibnamefont {Ho}},\ and\ \bibinfo {author}
  {\bibfnamefont {X.}~\bibnamefont {Rong}},\ }\bibfield  {title} {\bibinfo
  {title} {Observation of hierarchy of {Hilbert} space ergodicities in the
  quantum dynamics of a single spin},\ }\href
  {https://doi.org/10.1103/6msb-cxbc} {\bibfield  {journal} {\bibinfo
  {journal} {Phys. Rev. Lett.}\ }\textbf {\bibinfo {volume} {136}},\ \bibinfo
  {pages} {020401} (\bibinfo {year} {2026})}\BibitemShut {NoStop}%
\bibitem [{\citenamefont {Doherty}\ \emph {et~al.}(2013)\citenamefont
  {Doherty}, \citenamefont {Manson}, \citenamefont {Delaney}, \citenamefont
  {Jelezko}, \citenamefont {Wrachtrup},\ and\ \citenamefont
  {Hollenberg}}]{SM-Doherty2013}%
  \BibitemOpen
  \bibfield  {author} {\bibinfo {author} {\bibfnamefont {M.~W.}\ \bibnamefont
  {Doherty}}, \bibinfo {author} {\bibfnamefont {N.~B.}\ \bibnamefont {Manson}},
  \bibinfo {author} {\bibfnamefont {P.}~\bibnamefont {Delaney}}, \bibinfo
  {author} {\bibfnamefont {F.}~\bibnamefont {Jelezko}}, \bibinfo {author}
  {\bibfnamefont {J.}~\bibnamefont {Wrachtrup}},\ and\ \bibinfo {author}
  {\bibfnamefont {L.~C.~L.}\ \bibnamefont {Hollenberg}},\ }\bibfield  {title}
  {\bibinfo {title} {The nitrogen-vacancy colour centre in diamond},\ }\href
  {https://doi.org/10.1016/j.physrep.2013.02.001} {\bibfield  {journal}
  {\bibinfo  {journal} {Phys. Rep.}\ }\textbf {\bibinfo {volume} {528}},\
  \bibinfo {pages} {1} (\bibinfo {year} {2013})}\BibitemShut {NoStop}%
\end{thebibliography}

\clearpage
\onecolumngrid

\makeatletter
\let\addcontentsline\HSE@addcontentsline
\makeatother
\setcounter{secnumdepth}{3}
\setcounter{section}{0}
\setcounter{equation}{0}
\setcounter{figure}{0}
\setcounter{table}{0}
\renewcommand{\theequation}{S\arabic{equation}}
\renewcommand{\thefigure}{S\arabic{figure}}
\renewcommand{\thetable}{S\arabic{table}}
\renewcommand{\thesection}{S\arabic{section}}

\begin{center}
{\large\bfseries Supplemental Material for\\[2pt]
``Experimental Investigation of Tunable-Order Hilbert-Space Ergodicity''}
\end{center}

\tableofcontents

\medskip

\section{Theoretical Framework}
\label{sec:S1}

In this section we recall the definition of Hilbert-space ergodicity (HSE) used in the main text, recast it as a statement about a probability measure on the unitary group, reduce the $k$-HSE requirement to a set of conditions on the Fourier components of this measure, and establish the invariance of the limiting ensemble under a delayed start of the collection.
The measure formulation and the reduction to the symmetric subspace hold for a system of arbitrary dimension $d$, and the qubit structure enters only at the final step, where the operators on the symmetric subspace are decomposed into irreducible components.
These conditions, summarized in Eq.~\eqref{eq:S-final-condition}, are the starting point for the analysis of the $m$-tone drives in Sec.~\ref{sec:S2}.

\subsection{Measure-theoretic characterization of Hilbert-space ergodicity}
\label{sec:S1A}

In the main text, $k$-HSE is defined through the convergence of the temporal moments: the dynamics $U(t)$ satisfies $k$-HSE if, for every initial state $\rho_0 = |\psi(0)\rangle\langle\psi(0)|$, the trace distance $\Delta_T^{(k)}$ between the $k$th moment of the temporal ensemble,
\begin{equation}
    \rho_T^{(k)} = \frac{1}{T}\int_0^T \mathrm{d}t  \big[U(t) \rho_0 U^\dagger(t)\big]^{\otimes k},
    \label{eq:S-temporal-moment}
\end{equation}
and the corresponding Haar moment $\rho_{\mathrm{Haar}}^{(k)}$ vanishes as $T\to\infty$.
Equation~\eqref{eq:S-temporal-moment} averages the operator $U^{\otimes k}\rho_0^{\otimes k}U^{\dagger\otimes k}$ over the evolution operators visited up to time $T$, so the long-time dynamics of a $d$-dimensional system is naturally described by the probability measure that the trajectory $\{U(t)\}_{t\ge 0}$ induces on SU($d$)~\cite{SM-PilatowskyCameo2024}.
Throughout, averaging a function $f(U)$ with respect to a measure $\mu$ means
\begin{equation}
    \mathbb{E}_{U\sim\mu}\big[f(U)\big] = \int_{U\sim\mu} f(U) \mathrm{d}\mu .
    \label{eq:S-average-def}
\end{equation}

We call the measure induced by the trajectory the temporal measure $\nu$, defined through
\begin{equation}
    \mathbb{E}_{U\sim\nu}\big[f(U)\big]
    = \lim_{T\to\infty}\frac{1}{T}\int_0^T f\big(U(t)\big) \mathrm{d}t .
    \label{eq:S-nu-def}
\end{equation}
The reference measure of HSE is the Haar measure $\eta_{\mathrm{Haar}}$, the uniform probability measure on SU($d$).
Its defining property is the invariance under left and right multiplication by any fixed $V\in\mathrm{SU}(d)$,
\begin{equation}
    \mathbb{E}_{U\sim\eta_{\mathrm{Haar}}}\big[f(U)\big]
    = \mathbb{E}_{U\sim\eta_{\mathrm{Haar}}}\big[f(VU)\big]
    = \mathbb{E}_{U\sim\eta_{\mathrm{Haar}}}\big[f(UV)\big].
    \label{eq:S-haar-invariance}
\end{equation}

The finite-sample versions make the meanings of the two measures transparent.
For stroboscopic times $t = 0, 1, \ldots, T-1$, suppose the trajectory visits $N$ distinct operators $U_1, \ldots, U_N$ with multiplicities $C_1, \ldots, C_N$, so that each operator occurs with frequency $p_i = C_i/T$.
The time average of any function is then an average over these frequencies,
\begin{equation}
    \sum_{i=1}^{N} p_i f(U_i)
    = \frac{1}{T}\sum_{i=1}^{N} C_i f(U_i)
    = \frac{1}{T}\sum_{t=0}^{T-1} f\big(U(t)\big),
    \label{eq:S-discrete}
\end{equation}
and the temporal measure $\nu$ is the $T\to\infty$ limit of this frequency distribution, with Eq.~\eqref{eq:S-nu-def} as its continuous-time analogue. Similarly, if $N$ operators $V_1,\ldots,V_N$ are independently sampled from the Haar measure, their finite-sample average is
\begin{equation}
    \frac{1}{N}\sum_{i=1}^{N} f(V_i),
    \qquad
    V_i \sim \eta_{\mathrm{Haar}},
    \label{eq:S-haar-sample}
\end{equation}
which converges to $\mathbb{E}_{U\sim\eta_{\mathrm{Haar}}}[f(U)]$ as $N\to\infty$.
Thus the temporal measure describes the long-time frequency with which the evolution visits different operators, while the Haar measure describes the corresponding uniform distribution over SU($d$).

In this language, $k$-HSE takes the compact form
\begin{equation}
    \mathbb{E}_{U\sim\nu}\Big[U^{\otimes k}\rho_0^{\otimes k}U^{\dagger\otimes k}\Big]
    = \mathbb{E}_{U\sim\eta_{\mathrm{Haar}}}\Big[U^{\otimes k}\rho_0^{\otimes k}U^{\dagger\otimes k}\Big]
    \qquad \text{for every pure state } \rho_0 = |\psi(0)\rangle\langle\psi(0)| .
    \label{eq:S-kHSE-measure}
\end{equation}
By the invariance~\eqref{eq:S-haar-invariance}, the right-hand side of Eq.~\eqref{eq:S-kHSE-measure} is independent of $\rho_0$ and equals the Haar moment $\rho_{\mathrm{Haar}}^{(k)}$ of the main text.

\subsection{Reduction to the symmetric subspace}
\label{sec:S1B}

To analyze what Eq.~\eqref{eq:S-kHSE-measure} demands from $\nu$, we define the $k$th twirl induced by a measure $\mu$,
\begin{equation}
    \mathcal{T}^{(k)}_{\mu}(X) = \mathbb{E}_{U\sim\mu}\Big[U^{\otimes k} X  U^{\dagger\otimes k}\Big],
    \label{eq:S-twirl}
\end{equation}
a linear superoperator on the operators of $(\mathbb{C}^d)^{\otimes k}$.
The requirement~\eqref{eq:S-kHSE-measure} then reads
\begin{equation}
    \mathcal{T}^{(k)}_{\nu}\big(\rho_0^{\otimes k}\big) = \mathcal{T}^{(k)}_{\eta_{\mathrm{Haar}}}\big(\rho_0^{\otimes k}\big)
    \qquad \text{for every pure state } \rho_0 = |\psi(0)\rangle\langle\psi(0)| ,
    \label{eq:S-twirl-equality}
\end{equation}
where the equality is required only on $k$ copies of pure-state density operators, because the notion of HSE is intrinsically formulated for pure states.
Demanding the equality for all operators on $(\mathbb{C}^d)^{\otimes k}$ defines instead the stricter notion of $k$-unitary ergodicity~\cite{SM-PilatowskyCameo2024}, which implies $k$-HSE but not conversely, and is not the subject of this work.

These inputs $\rho_0^{\otimes k} = \big(|\psi(0)\rangle\langle\psi(0)|\big)^{\otimes k}$ live in a small subspace.
The symmetric subspace of $(\mathbb{C}^d)^{\otimes k}$ is
\begin{equation}
    \mathrm{Sym}^k(\mathbb{C}^d) = \mathrm{span}\Big\{ |\Psi\rangle \in (\mathbb{C}^d)^{\otimes k} : P_\pi |\Psi\rangle = |\Psi\rangle \ \ \forall \pi \in S_k \Big\},
    \label{eq:S-sym-def}
\end{equation}
where $S_k$ is the group of permutations of $k$ objects and $P_\pi$ permutes the $k$ copies according to $\pi$, and the projector onto it reads~\cite{SM-Harrow2013}
\begin{equation}
    \Pi_{\mathrm{sym}} = \frac{1}{k!}\sum_{\pi\in S_k} P_\pi .
    \label{eq:S-sym-projector}
\end{equation}
Every $k$-copy pure state lies in this subspace,
\begin{equation}
    |\psi(0)\rangle^{\otimes k} \in \mathrm{Sym}^k(\mathbb{C}^d),
    \qquad\text{equivalently}\qquad
    \rho_0^{\otimes k} = \Pi_{\mathrm{sym}} \rho_0^{\otimes k} \Pi_{\mathrm{sym}},
    \label{eq:S-pure-in-sym}
\end{equation}
and every tensor power commutes with the projector,
\begin{equation}
    U^{\otimes k} \Pi_{\mathrm{sym}} = \Pi_{\mathrm{sym}}  U^{\otimes k} .
    \label{eq:S-commute}
\end{equation}
Combining the two relations, for any measure $\mu$,
\begin{equation}
\begin{aligned}
    \mathcal{T}^{(k)}_{\mu}\big(\rho_0^{\otimes k}\big)
    &= \int_{U\sim\mu} U^{\otimes k} \rho_0^{\otimes k} U^{\dagger\otimes k} \mathrm{d}\mu \\
    &= \int_{U\sim\mu} U^{\otimes k} \Pi_{\mathrm{sym}} \rho_0^{\otimes k} \Pi_{\mathrm{sym}} U^{\dagger\otimes k} \mathrm{d}\mu \\
    &= \int_{U\sim\mu} \Pi_{\mathrm{sym}} U^{\otimes k} \rho_0^{\otimes k} U^{\dagger\otimes k} \Pi_{\mathrm{sym}} \mathrm{d}\mu \\
    &= \Pi_{\mathrm{sym}}\left(\int_{U\sim\mu} U^{\otimes k} \rho_0^{\otimes k} U^{\dagger\otimes k} \mathrm{d}\mu\right)\Pi_{\mathrm{sym}} \\
    &= \Pi_{\mathrm{sym}}  \mathcal{T}^{(k)}_{\mu}\big(\rho_0^{\otimes k}\big)  \Pi_{\mathrm{sym}} ,
\end{aligned}
    \label{eq:S-twirl-sym}
\end{equation}
so both the inputs $\rho_0^{\otimes k}$ and the outputs $\mathcal{T}^{(k)}_{\mu}\big(\rho_0^{\otimes k}\big)$ of the two twirls $\mathcal{T}^{(k)}_{\nu}$ and $\mathcal{T}^{(k)}_{\eta_{\mathrm{Haar}}}$ lie entirely in $\mathrm{Sym}^k(\mathbb{C}^d)$, whose dimension is $\binom{d+k-1}{k}$.
The $k$-HSE requirement therefore only concerns the action of the twirls on this subspace, with no requirement on the other subspaces.
Up to this point, every statement holds for an arbitrary dimension $d$.

\subsection{Decomposition of the qubit operator space}
\label{sec:S1C}

To carry out the decomposition explicitly, we now specialize to the qubit case, $d=2$, on which the main text focuses.
The goal of this subsection is to restrict the $k$-HSE requirement to the symmetric subspace and to decompose it into sectors, reducing it to the matching of the first $k$ Fourier components of the temporal and Haar measures.

On the symmetric subspace, the tensor power induces a representation of SU(2),
\begin{equation}
    D_k(U) = U^{\otimes k}\big|_{\mathrm{Sym}^k(\mathbb{C}^2)} ,
    \label{eq:S-Dk}
\end{equation}
that is, $D_k(U)$ is $U^{\otimes k}$ with its domain restricted to $\mathrm{Sym}^k(\mathbb{C}^2)$, well defined as an operator on this subspace because Eq.~\eqref{eq:S-commute} guarantees that the image also stays in $\mathrm{Sym}^k(\mathbb{C}^2)$.
The carrier space has dimension
\begin{equation}
    \dim \mathrm{Sym}^k(\mathbb{C}^2) = \binom{k+1}{k} = k+1 .
    \label{eq:S-sym-dim}
\end{equation}
This representation is irreducible~\cite{SM-Harrow2013}.
Since SU(2) possesses exactly one irreducible representation of each dimension up to unitary equivalence, the dimension $k+1$ identifies $D_k$ with the standard spin-$j$ irreducible representation: in the angular-momentum basis $\{|j,m\rangle\}_{m=-j}^{j}$ of the symmetric subspace,
\begin{equation}
    D_k(U) = D^{(j)}(U), \qquad j = \frac{k}{2} ,
    \label{eq:S-Dk-Dj}
\end{equation}
with $D^{(j)}$ the spin-$j$ Wigner rotation matrix.
We write $V_j = \mathrm{span}\big(\{|j,m\rangle\}_{m=-j}^{j}\big)$ with $j=k/2$ for its carrier space.

By Eq.~\eqref{eq:S-twirl-sym}, the inputs $\rho_0^{\otimes k}$ and the outputs $\mathcal{T}^{(k)}_{\mu}\big(\rho_0^{\otimes k}\big)$ of the two twirls in the $k$-HSE requirement~\eqref{eq:S-twirl-equality} are operators on $V_j$ with $j = k/2$, that is, elements of $\mathrm{Hom}(V_j, V_j)$, the space of linear maps from $V_j$ to itself.
Since the $k$-copy pure-state projectors span $\mathrm{Hom}(V_j, V_j)$~\cite{SM-Harrow2013}, the $k$-HSE requirement~\eqref{eq:S-twirl-equality}, imposed for every pure initial state $\rho_0$, is equivalent to the same equality with $\rho_0^{\otimes k}$ replaced by an arbitrary operator $X$,
\begin{equation}
    \int_{U\sim\nu} D^{(j)}(U)  X  D^{(j)}(U)^\dagger \mathrm{d}\nu
    = \int_{U\sim\eta_{\mathrm{Haar}}} D^{(j)}(U)  X  D^{(j)}(U)^\dagger \mathrm{d}\eta_{\mathrm{Haar}}
    \qquad \forall X \in \mathrm{Hom}(V_j, V_j) .
    \label{eq:S-twirl-on-Vj}
\end{equation}
To decompose this condition into independent pieces, we use the following structure of the SU(2) representations.
For finite-dimensional spaces $V$ and $W$,
\begin{equation}
    \mathrm{Hom}(W, V)  \cong  V \otimes W^* ,
    \label{eq:S-hom-iso}
\end{equation}
where $W^*$ is the dual space and $\mathrm{Hom}(W,V)$ the space of linear maps from $W$ to $V$: in Dirac notation, with orthonormal bases $\{|v_i\rangle\}$ of $V$ and $\{|w_j\rangle\}$ of $W$, any element of $V\otimes W^*$ takes the form $\sum_{ij} c_{ij}  |v_i\rangle\langle w_j|$, which is precisely a linear map from $W$ to $V$.
Moreover, every irreducible representation of SU(2) is self-dual,
\begin{equation}
    V_j^*  \cong  V_j ,
    \label{eq:S-selfdual}
\end{equation}
so that
\begin{equation}
    \mathrm{Hom}(V_j, V_j)  \cong  V_j \otimes V_j^*  \cong  V_j \otimes V_j .
    \label{eq:S-homVV}
\end{equation}
The Clebsch--Gordan theorem for SU(2) decomposes the tensor product of two irreducible representations as
\begin{equation}
    V_{j_1} \otimes V_{j_2}  \cong  \bigoplus_{J=|j_1-j_2|}^{j_1+j_2} V_J ,
    \label{eq:S-CG-general}
\end{equation}
with $J$ increasing in integer steps.
Substituting $j_1 = j_2 = j = k/2$ into Eqs.~\eqref{eq:S-homVV} and \eqref{eq:S-CG-general} yields
\begin{equation}
    \mathrm{Hom}(V_j, V_j)  \cong  \bigoplus_{\ell=0}^{k} V_\ell ,
    \label{eq:S-CG}
\end{equation}
in complete analogy with the addition of two angular momenta $j = k/2$.

The decomposition~\eqref{eq:S-CG} provides a basis of $\mathrm{Hom}(V_j, V_j)$ consisting of the spherical tensor operators $\{T^{(\ell)}_m\}$ with $\ell = 0, 1, \ldots, k$ and $m = -\ell, \ldots, \ell$, transforming as
\begin{equation}
    D^{(j)}(U)  T^{(\ell)}_m  D^{(j)}(U)^\dagger = \sum_{m'=-\ell}^{\ell} D^{(\ell)}_{m'm}(U)  T^{(\ell)}_{m'} ,
    \label{eq:S-tensor-transform}
\end{equation}
where $D^{(\ell)}_{m'm}(U)$ are the matrix elements of the spin-$\ell$ irreducible representation.
We refer to the $(2\ell{+}1)$-dimensional operator subspace spanned by $\{T^{(\ell)}_m\}_{m=-\ell}^{\ell}$ at fixed $\ell$, the $\ell$th summand of Eq.~\eqref{eq:S-CG}, as the sector $\ell$.
Any $X \in \mathrm{Hom}(V_j,V_j)$ can then be uniquely expanded as
\begin{equation}
    X = \sum_{\ell=0}^{k}\sum_{m=-\ell}^{\ell} x_{\ell m}  T^{(\ell)}_m .
    \label{eq:S-expansion}
\end{equation}
The basis can be chosen orthonormal with respect to the Hilbert--Schmidt inner product $\langle A, B \rangle = \mathrm{tr}[A^\dagger B]$, a normalization we adopt throughout, so that the coefficients are read off as $x_{\ell m} = \mathrm{tr}\big[T^{(\ell)\dagger}_m X\big]$.
Substituting Eq.~\eqref{eq:S-expansion} into Eq.~\eqref{eq:S-twirl-on-Vj} and applying the transformation law~\eqref{eq:S-tensor-transform} to each term,
\begin{equation}
    \sum_{\ell, m, m'} x_{\ell m} \left(\int_{U\sim\nu} D^{(\ell)}_{m'm}(U) \mathrm{d}\nu\right) T^{(\ell)}_{m'}
    = \sum_{\ell, m, m'} x_{\ell m} \left(\int_{U\sim\eta_{\mathrm{Haar}}} D^{(\ell)}_{m'm}(U) \mathrm{d}\eta_{\mathrm{Haar}}\right) T^{(\ell)}_{m'} .
    \label{eq:S-substituted}
\end{equation}
Since Eq.~\eqref{eq:S-substituted} holds for arbitrary coefficients $x_{\ell m}$ and the operators $T^{(\ell)}_{m'}$ are linearly independent, the averaged representation matrices must agree component by component,
\begin{equation}
    \int_{U\sim\nu} D^{(\ell)}(U) \mathrm{d}\nu
    = \int_{U\sim\eta_{\mathrm{Haar}}} D^{(\ell)}(U) \mathrm{d}\eta_{\mathrm{Haar}},
    \qquad \ell = 0, 1, \ldots, k .
    \label{eq:S-matrix-equality}
\end{equation}
It is convenient to regard such averages as the Fourier components of a measure,
\begin{equation}
    \hat{\mu}(\ell) = \int_{U\sim\mu} D^{(\ell)}(U) \mathrm{d}\mu ,
    \label{eq:S-fourier}
\end{equation}
in terms of which Eq.~\eqref{eq:S-matrix-equality} reads
\begin{equation}
    \hat{\nu}(\ell) = \hat{\eta}_{\mathrm{Haar}}(\ell),
    \qquad \ell = 0, 1, \ldots, k .
    \label{eq:S-fourier-equality}
\end{equation}

\subsection{Conditions on the temporal measure}
\label{sec:S1D}

This subsection evaluates the Haar side of Eq.~\eqref{eq:S-fourier-equality} once and for all, reducing the $k$-HSE requirement to the vanishing of the first $k$ Fourier components of the temporal measure.
Left-invariance of the Haar measure, together with the representation property $D^{(\ell)}(VU) = D^{(\ell)}(V) D^{(\ell)}(U)$, gives
\begin{equation}
    \hat{\eta}_{\mathrm{Haar}}(\ell)
    = \int_{U\sim\eta_{\mathrm{Haar}}} D^{(\ell)}(VU) \mathrm{d}\eta_{\mathrm{Haar}}
    = D^{(\ell)}(V) \int_{U\sim\eta_{\mathrm{Haar}}} D^{(\ell)}(U) \mathrm{d}\eta_{\mathrm{Haar}}
    = D^{(\ell)}(V)  \hat{\eta}_{\mathrm{Haar}}(\ell) ,
    \label{eq:S-left-inv}
\end{equation}
and right-invariance gives, in the same way,
\begin{equation}
    \hat{\eta}_{\mathrm{Haar}}(\ell)
    = \int_{U\sim\eta_{\mathrm{Haar}}} D^{(\ell)}(UV) \mathrm{d}\eta_{\mathrm{Haar}}
    = \hat{\eta}_{\mathrm{Haar}}(\ell)  D^{(\ell)}(V) .
    \label{eq:S-right-inv}
\end{equation}
Hence $\hat{\eta}_{\mathrm{Haar}}(\ell)$ commutes with the whole irreducible representation,
\begin{equation}
    \hat{\eta}_{\mathrm{Haar}}(\ell) = D^{(\ell)}(V)  \hat{\eta}_{\mathrm{Haar}}(\ell) = \hat{\eta}_{\mathrm{Haar}}(\ell)  D^{(\ell)}(V)
    \qquad \forall V \in \mathrm{SU}(2),
    \label{eq:S-haar-commute}
\end{equation}
and Schur's lemma~\cite{SM-Harrow2013} forces it to be proportional to the identity matrix $I_{2\ell+1}$ of dimension $2\ell+1$,
\begin{equation}
    \hat{\eta}_{\mathrm{Haar}}(\ell) = c_\ell  I_{2\ell+1} .
    \label{eq:S-schur}
\end{equation}
To determine $c_\ell$, introduce the character of the irreducible representation,
\begin{equation}
    \chi_\ell(U) = \mathrm{tr}\big[D^{(\ell)}(U)\big],
    \label{eq:S-character}
\end{equation}
which obeys the Schur orthogonality relation
\begin{equation}
    \int_{U\sim\eta_{\mathrm{Haar}}} \chi_{\ell'}(U)^* \chi_{\ell}(U) \mathrm{d}\eta_{\mathrm{Haar}} = \delta_{\ell'\ell} .
    \label{eq:S-orthogonality}
\end{equation}
For every nontrivial irreducible representation, the character is orthogonal to the trivial character $\chi_0(U) \equiv 1$, so
\begin{equation}
    \int_{U\sim\eta_{\mathrm{Haar}}} \chi_{\ell}(U) \mathrm{d}\eta_{\mathrm{Haar}} = 0,
    \qquad \ell \ge 1 .
    \label{eq:S-character-zero}
\end{equation}
Taking the trace of Eq.~\eqref{eq:S-schur} and using Eq.~\eqref{eq:S-character-zero},
\begin{equation}
    \mathrm{tr}\big[\hat{\eta}_{\mathrm{Haar}}(\ell)\big] = c_\ell (2\ell+1)
    = \int_{U\sim\eta_{\mathrm{Haar}}} \chi_\ell(U) \mathrm{d}\eta_{\mathrm{Haar}} ,
    \label{eq:S-trace-chain}
\end{equation}
which fixes $c_\ell = 0$ for all $\ell \ge 1$, while $D^{(0)}(U) \equiv 1$ directly gives $c_0 = 1$.
Therefore,
\begin{equation}
    \hat{\eta}_{\mathrm{Haar}}(\ell) =
    \begin{cases}
        1, & \ell = 0,\\[2pt]
        0_{(2\ell+1)\times(2\ell+1)}, & \ell \ge 1 .
    \end{cases}
    \label{eq:S-haar-fourier}
\end{equation}

Combining Eqs.~\eqref{eq:S-matrix-equality} and \eqref{eq:S-haar-fourier}, the $k$-HSE requirement becomes
\begin{equation}
    \hat{\nu}(\ell) = \int_{U\sim\nu} D^{(\ell)}(U) \mathrm{d}\nu =
    \begin{cases}
        1, & \ell = 0,\\[2pt]
        0_{(2\ell+1)\times(2\ell+1)}, & \ell \ge 1 ,
    \end{cases}
    \qquad \ell = 0, 1, \ldots, k .
    \label{eq:S-combined-condition}
\end{equation}
The $\ell = 0$ condition holds automatically, because $D^{(0)}(U) \equiv 1$ and every probability measure is normalized.
The $k$-HSE requirement finally reduces to the vanishing of the first $k$ Fourier components of the temporal measure,
\begin{equation}
    \hat{\nu}(\ell) = \int_{U\sim\nu} D^{(\ell)}(U) \mathrm{d}\nu = 0_{(2\ell+1)\times(2\ell+1)},
    \qquad \ell = 1, 2, \ldots, k .
    \label{eq:S-final-condition}
\end{equation}
Equation~\eqref{eq:S-final-condition} also makes the hierarchy of HSE manifest, as the conditions for $(k{+}1)$-HSE contain those for $k$-HSE.
In Sec.~\ref{sec:S2} we evaluate $\hat{\nu}(\ell)$ for the $m$-tone drives and turn Eq.~\eqref{eq:S-final-condition} into the Legendre-polynomial conditions quoted in the main text.

\subsection{Invariance properties of the infinite-time ensemble}
\label{sec:S1E}

The main text states that initial-state information survives in the late-time statistics of the temporal ensemble.
For this statement to be operationally meaningful, reading out the information must not require access to the early evolution: an observer who starts collecting at $t=0$ holds the initial state itself, and extracting its information from the ensemble would be an empty exercise.
This subsection shows that the readout indeed needs no early history.

Let the temporal moment be collected only from time $t_0$ onward,
\begin{equation}
    \rho_{T, t_0}^{(k)} = \frac{1}{T-t_0}\int_{t_0}^{T} F(t) \mathrm{d}t ,
    \qquad
    F(t) = \big[U(t) \rho_0 U^\dagger(t)\big]^{\otimes k},
    \label{eq:S-shifted-moment}
\end{equation}
so that $\rho_{T,0}^{(k)}$ recovers Eq.~\eqref{eq:S-temporal-moment}.
Splitting the full-window integral as $\int_0^T = \int_0^{t_0} + \int_{t_0}^{T}$, the two moments are related by
\begin{equation}
\begin{aligned}
    \rho_{T, t_0}^{(k)} - \rho_{T, 0}^{(k)}
    &= \frac{1}{T-t_0}\int_{t_0}^{T} F(t) \mathrm{d}t
     - \frac{1}{T}\int_{0}^{t_0} F(t) \mathrm{d}t
     - \frac{1}{T}\int_{t_0}^{T} F(t) \mathrm{d}t \\
    &= \left(\frac{1}{T-t_0} - \frac{1}{T}\right)\int_{t_0}^{T} F(t) \mathrm{d}t
     - \frac{1}{T}\int_{0}^{t_0} F(t) \mathrm{d}t \\
    &= \frac{t_0}{T (T-t_0)}\int_{t_0}^{T} F(t) \mathrm{d}t
     - \frac{1}{T}\int_{0}^{t_0} F(t) \mathrm{d}t .
\end{aligned}
    \label{eq:S-window-difference}
\end{equation}
Each $F(t)$ is a rank-one projector with unit trace norm, $\|F(t)\|_1 = 1$, where $\|X\|_1 = \mathrm{tr}\sqrt{X^\dagger X}$ denotes the trace norm entering the trace distance of the main text.
Applying the triangle inequality to the two terms of Eq.~\eqref{eq:S-window-difference} and moving the norm inside the integrals,
\begin{equation}
\begin{aligned}
    \frac{1}{2}\Big\| \rho_{T, t_0}^{(k)} - \rho_{T, 0}^{(k)} \Big\|_1
    &\le \frac{1}{2}\left[ \frac{t_0}{T (T-t_0)} \left\| \int_{t_0}^{T} F(t) \mathrm{d}t \right\|_1
        + \frac{1}{T} \left\| \int_{0}^{t_0} F(t) \mathrm{d}t \right\|_1 \right] \\
    &\le \frac{1}{2}\left[ \frac{t_0}{T (T-t_0)} \int_{t_0}^{T} \big\|F(t)\big\|_1 \mathrm{d}t
        + \frac{1}{T} \int_{0}^{t_0} \big\|F(t)\big\|_1 \mathrm{d}t \right] \\
    &= \frac{1}{2}\left[ \frac{t_0}{T (T-t_0)} (T-t_0) + \frac{t_0}{T} \right]
     = \frac{t_0}{T}  \xrightarrow[T\to\infty]{}  0 .
\end{aligned}
    \label{eq:S-window-bound}
\end{equation}
Writing $\Delta_{T,t_0}^{(k)}$ for the trace distance evaluated with the moment~\eqref{eq:S-shifted-moment}, the triangle inequality gives
\begin{equation}
    \Big| \Delta_{T, t_0}^{(k)} - \Delta_{T, 0}^{(k)} \Big|
     \le  \frac{1}{2}\Big\| \rho_{T, t_0}^{(k)} - \rho_{T, 0}^{(k)} \Big\|_1
     \le  \frac{t_0}{T} ,
    \label{eq:S-window-triangle}
\end{equation}
so the limit of the trace distance exists for the delayed collection exactly when it exists for the full one, and the two limits coincide.
Whenever the limits exist we write $\rho_\infty^{(k)} = \lim_{T\to\infty}\rho_T^{(k)}$ and $\Delta_\infty^{(k)} = \lim_{T\to\infty}\Delta_T^{(k)}$, the notation used in the main text and in the remainder of this Supplemental Material.
An observer who knows the drive but joins the experiment only at time $t_0$ therefore obtains the same limiting statistics, and with them the same fingerprint of the initial state.
The same bound also shows that the experimental plateau is robust against discarding early data.

\section{The $m$-Tone Drives and the $k$-HSE Conditions}
\label{sec:S2}

In this section we evaluate the Fourier components $\hat{\nu}(\ell)$ for the $m$-tone drives, prove the Legendre-polynomial conditions quoted as Eq.~(3) of the main text, establish the maximal achievable order $k = 2m{-}3$, show the invariance of the achievable order under permutations of the inter-axis angles, and present additional drive examples.

\subsection{From the $m$-tone drive to the Legendre conditions}
\label{sec:S2A}

The $m$-tone drive consists of successive single-axis rotations at $m$ rationally independent frequencies,
\begin{equation}
    U(t) = e^{-i\omega_m t H_m}\cdots e^{-i\omega_2 t H_2}  e^{-i\omega_1 t H_1}
    = \prod_{j=m}^{1} e^{-i\omega_j t H_j},
    \qquad H_j = \boldsymbol{n}_j\cdot\boldsymbol{\sigma},
    \label{eq:S-mtone}
\end{equation}
where the unit vectors $\{\boldsymbol{n}_j\}$ set the rotation axes and $\boldsymbol{\sigma} = (\sigma_x, \sigma_y, \sigma_z)$ collects the Pauli matrices.
Collecting the phases into
\begin{equation}
    \boldsymbol{\theta}(t) = \big(\theta_1, \theta_2, \ldots, \theta_m\big) = \big(\omega_1 t,  \omega_2 t,  \ldots,  \omega_m t\big) \bmod 2\pi ,
    \label{eq:S-torus-point}
\end{equation}
the evolution operator is represented by a point on the $m$-torus $\mathbb{T}^m$,
\begin{equation}
    U(t) \equiv U(\boldsymbol{\theta}) = \prod_{j=m}^{1} g_j(\theta_j),
    \qquad g_j(\theta_j) = e^{-i\theta_j H_j} .
    \label{eq:S-U-theta}
\end{equation}
For rationally independent $\{\omega_j\}$, Weyl's equidistribution theorem guarantees that $\boldsymbol{\theta}(t)$ becomes uniformly distributed on $\mathbb{T}^m$~\cite{SM-Weyl1916}, so the temporal average of Eq.~\eqref{eq:S-nu-def} turns into a uniform torus average,
\begin{equation}
    \mathbb{E}_{U\sim\nu}\big[f(U)\big]
    = \int_{\boldsymbol{\theta}\in\mathbb{T}^m} f\big(U(\boldsymbol{\theta})\big) \frac{\mathrm{d}^m\theta}{(2\pi)^m}
    = \int_{\theta_m\sim\mu_m}\cdots\int_{\theta_1\sim\mu_1} f\Big(\prod_{j=m}^{1} g_j(\theta_j)\Big) \mathrm{d}\mu_1\cdots\mathrm{d}\mu_m ,
    \label{eq:S-torus-average}
\end{equation}
where $\mu_j = \mathrm{d}\theta_j/2\pi$ denotes the uniform measure on the circle traversed by the $j$th phase.
Introducing the convolution of two probability measures on the group,
\begin{equation}
    \int_{W\sim\mu_a*\mu_b} f(W) \mathrm{d}(\mu_a*\mu_b)
    = \int_{U_a\sim\mu_a}\int_{U_b\sim\mu_b} f(U_a U_b) \mathrm{d}\mu_a \mathrm{d}\mu_b ,
    \label{eq:S-convolution-def}
\end{equation}
the nested structure of Eq.~\eqref{eq:S-torus-average} states that the temporal measure of the $m$-tone drive is an $m$-fold convolution,
\begin{equation}
    \nu = \mu_m * \mu_{m-1} * \cdots * \mu_1 .
    \label{eq:S-nu-convolution}
\end{equation}

Convolutions become products in Fourier space.
Using the representation property $D^{(\ell)}(U_a U_b) = D^{(\ell)}(U_a) D^{(\ell)}(U_b)$,
\begin{equation}
    \widehat{\mu_a*\mu_b}(\ell)
    = \int_{W\sim\mu_a*\mu_b} D^{(\ell)}(W) \mathrm{d}(\mu_a*\mu_b)
    = \int_{U_a\sim\mu_a}\int_{U_b\sim\mu_b} D^{(\ell)}(U_a) D^{(\ell)}(U_b) \mathrm{d}\mu_a \mathrm{d}\mu_b
    = \hat{\mu}_a(\ell) \hat{\mu}_b(\ell) ,
    \label{eq:S-fourier-conv}
\end{equation}
so that Eq.~\eqref{eq:S-nu-convolution} yields
\begin{equation}
    \hat{\nu}(\ell) = \hat{\mu}_m(\ell) \hat{\mu}_{m-1}(\ell)\cdots\hat{\mu}_1(\ell) .
    \label{eq:S-nu-product}
\end{equation}

It remains to compute the single-tone factors
\begin{equation}
    \hat{\mu}_j(\ell) = \int_{\theta_j\sim\mu_j} D^{(\ell)}\big(g_j(\theta_j)\big) \mathrm{d}\mu_j
    = \int_0^{2\pi} D^{(\ell)}\big(e^{-i\theta_j \boldsymbol{n}_j\cdot\boldsymbol{\sigma}}\big) \frac{\mathrm{d}\theta_j}{2\pi} .
    \label{eq:S-single-factor}
\end{equation}
Choose $R_j \in \mathrm{SU}(2)$ that rotates the $z$ axis to the $\boldsymbol{n}_j$ axis, $R_j \sigma_z R_j^\dagger = \boldsymbol{n}_j\cdot\boldsymbol{\sigma}$, so that
\begin{equation}
    g_j(\theta_j) = e^{-i\theta_j \boldsymbol{n}_j\cdot\boldsymbol{\sigma}} = R_j  e^{-i\theta_j\sigma_z}  R_j^\dagger ,
    \label{eq:S-rotate-axis}
\end{equation}
and by the representation property,
\begin{equation}
    \hat{\mu}_j(\ell)
    = D^{(\ell)}(R_j)\left[\int_0^{2\pi} D^{(\ell)}\big(e^{-i\theta_j\sigma_z}\big) \frac{\mathrm{d}\theta_j}{2\pi}\right] D^{(\ell)}(R_j)^\dagger .
    \label{eq:S-factor-rotated}
\end{equation}
The circle average is computed in the eigenbasis $\{|\ell, m\rangle : m = -\ell, \ldots, \ell\}$ of $J_z^{(\ell)}$, the $z$ component of the angular momentum in the spin-$\ell$ representation.
Since $\sigma_z = 2 J_z^{(1/2)}$, the group element $e^{-i\theta_j\sigma_z}$ is represented in the spin-$\ell$ representation as $e^{-i2\theta_j J_z^{(\ell)}}$, so
\begin{equation}
    D^{(\ell)}\big(e^{-i\theta_j\sigma_z}\big) |\ell, m\rangle = e^{-i2m\theta_j} |\ell, m\rangle ,
    \label{eq:S-phase-action}
\end{equation}
and the average over one full circle retains only the zero-weight component,
\begin{equation}
    \int_0^{2\pi} e^{-i2m\theta_j} \frac{\mathrm{d}\theta_j}{2\pi} = \delta_{m0} .
    \label{eq:S-delta-m0}
\end{equation}
Expanding the average in this basis,
\begin{equation}
    \int_0^{2\pi} D^{(\ell)}\big(e^{-i\theta_j\sigma_z}\big) \frac{\mathrm{d}\theta_j}{2\pi}
    = \sum_{m, m'} |\ell, m\rangle\langle \ell, m|\left[\int_0^{2\pi} D^{(\ell)}\big(e^{-i\theta_j\sigma_z}\big) \frac{\mathrm{d}\theta_j}{2\pi}\right]|\ell, m'\rangle\langle \ell, m'|
    = \sum_{m, m'} \delta_{m'0} \delta_{mm'}  |\ell, m\rangle\langle \ell, m'| ,
    \label{eq:S-basis-expansion}
\end{equation}
which gives
\begin{equation}
    \int_0^{2\pi} D^{(\ell)}\big(e^{-i\theta_j\sigma_z}\big) \frac{\mathrm{d}\theta_j}{2\pi}
    =
    \begin{cases}
        |\ell, 0\rangle\langle \ell, 0| , & \ell = 0, 1, 2, \ldots \\[2pt]
        0_{(2\ell+1)\times(2\ell+1)} , & \ell = \tfrac12, \tfrac32, \tfrac52, \ldots
    \end{cases}
    \label{eq:S-circle-average}
\end{equation}
Only integer $\ell$ occurs in the HSE conditions, so with $|\ell, 0\rangle_{\boldsymbol{n}_j} = D^{(\ell)}(R_j) |\ell, 0\rangle$ denoting the zero-weight eigenstate of $J_{\boldsymbol{n}_j}$ along the direction $\boldsymbol{n}_j$,
\begin{equation}
    \hat{\mu}_j(\ell) = |\ell, 0\rangle_{\boldsymbol{n}_j}~{}_{\boldsymbol{n}_j}\langle \ell, 0| .
    \label{eq:S-factor-final}
\end{equation}
Each single-tone factor is a rank-one projector onto the zero-weight state along its own axis.

Putting Eqs.~\eqref{eq:S-nu-product} and \eqref{eq:S-factor-final} together, the product of projectors telescopes,
\begin{equation}
    \hat{\nu}(\ell) = \prod_{j=m}^{1} |\ell, 0\rangle_{\boldsymbol{n}_j}~{}_{\boldsymbol{n}_j}\langle \ell, 0|
    = \left[\prod_{j=1}^{m-1} {}_{\boldsymbol{n}_{j+1}}\langle \ell, 0|\ell, 0\rangle_{\boldsymbol{n}_j}\right] |\ell, 0\rangle_{\boldsymbol{n}_m}~{}_{\boldsymbol{n}_1}\langle \ell, 0| .
    \label{eq:S-telescope}
\end{equation}
The overlap of zero-weight states along two axes separated by the angle $\phi_j = \angle(\boldsymbol{n}_j, \boldsymbol{n}_{j+1})$ is the Wigner small-$d$ matrix element $d^{\ell}_{00}$, which coincides with the Legendre polynomial,
\begin{equation}
    {}_{\boldsymbol{n}_{j+1}}\langle \ell, 0|\ell, 0\rangle_{\boldsymbol{n}_j} = d^{\ell}_{00}(\phi_j) = P_\ell(\cos\phi_j) ,
    \label{eq:S-legendre-overlap}
\end{equation}
so that finally
\begin{equation}
    \hat{\nu}(\ell) = \left[\prod_{j=1}^{m-1} P_\ell(\cos\phi_j)\right] |\ell, 0\rangle_{\boldsymbol{n}_m}~{}_{\boldsymbol{n}_1}\langle \ell, 0| .
    \label{eq:S-nu-final}
\end{equation}
Since the rank-one operator $|\ell, 0\rangle_{\boldsymbol{n}_m}~{}_{\boldsymbol{n}_1}\langle \ell, 0|$ never vanishes, combining Eq.~\eqref{eq:S-nu-final} with the requirement~\eqref{eq:S-final-condition} shows that the $m$-tone drive achieves $k$-HSE exactly when
\begin{equation}
    \prod_{j=1}^{m-1} P_\ell(\cos\phi_j) = 0,
    \qquad \ell = 1, 2, \ldots, k ,
    \label{eq:S-legendre-conditions}
\end{equation}
which is Eq.~(3) of the main text.

\subsection{Maximal order $k = 2m{-}3$}
\label{sec:S2B}

Two properties of Legendre polynomials determine how efficiently the conditions~\eqref{eq:S-legendre-conditions} can be satisfied.

\emph{Property 1.}---$P_\ell(0) = 0$ if and only if $\ell$ is odd.
This follows from the parity $P_\ell(-x) = (-1)^\ell P_\ell(x)$: odd-degree polynomials are odd functions and vanish at the origin, while even-degree ones take nonzero values there.

\emph{Property 2.}---The Legendre polynomials obey the recurrence relation
\begin{equation}
    (\ell+1)  P_{\ell+1}(x) = (2\ell+1)  x  P_\ell(x) - \ell  P_{\ell-1}(x) .
    \label{eq:S-recurrence}
\end{equation}

The recurrence relation of Property 2 implies the following lemma:  For $x \neq 0$, no two Legendre polynomials of degrees $\ell$ and $\ell{+}2$ vanish simultaneously at $x$.
To prove this, suppose $P_\ell(x) = 0$ and $P_{\ell+2}(x) = 0$ with $x \neq 0$.
Applying Eq.~\eqref{eq:S-recurrence} with $\ell \to \ell{+}1$ and using $P_\ell(x) = 0$,
\begin{equation}
    (\ell+2)  P_{\ell+2}(x) = (2\ell+3)  x  P_{\ell+1}(x) ,
    \label{eq:S-lemma-step1}
\end{equation}
so $P_{\ell+2}(x) = 0$ and $x \neq 0$ force $P_{\ell+1}(x) = 0$.
Applying Eq.~\eqref{eq:S-recurrence} at $\ell$ with $P_\ell(x) = P_{\ell+1}(x) = 0$ then gives
\begin{equation}
    0 = (\ell+1)  P_{\ell+1}(x) = (2\ell+1)  x  P_\ell(x) - \ell  P_{\ell-1}(x) = -\ell  P_{\ell-1}(x) ,
    \label{eq:S-lemma-step2}
\end{equation}
so $P_{\ell-1}(x) = 0$ as well.
The pair of vanishing polynomials of degrees $(\ell, \ell{+}2)$ has thus produced a vanishing pair of degrees $(\ell{-}1, \ell{+}1)$, and repeating the argument descends step by step until it reaches $P_0(x) = 0$, contradicting $P_0(x) = 1$.

The lemma implies that among the even degrees $\ell = 2, 4, 6, \ldots$, a fixed angle $\phi_j$ with $\cos\phi_j \neq 0$ can satisfy $P_\ell(\cos\phi_j) = 0$ for at most one even $\ell$, so each even-degree condition consumes an angle of its own.
The optimization of the $m{-}1$ angles then proceeds as follows.
The condition at $\ell = 1$ reads $P_1(\cos\phi_j) = \cos\phi_j = 0$ and forces at least one angle to equal $\pi/2$, the only zero of $\cos\phi_j$ on $[0, \pi]$; by Property 1, this single angle then satisfies all odd-degree conditions simultaneously.
Each of the remaining $m{-}2$ angles satisfies at most one even-degree condition, so the consecutive conditions $\ell = 1, 2, \ldots, k$ can be met at best when the $m{-}2$ angles cover the even degrees $2, 4, \ldots, 2(m{-}2)$, one each.

The hierarchy of HSE makes the resulting pattern transparent: by Eq.~\eqref{eq:S-legendre-conditions}, the drive achieves $k$-HSE exactly when the conditions hold consecutively for $\ell = 1, \ldots, k$, so the first degree whose product fails to vanish caps the achievable order at its predecessor:
\begin{itemize}
    \item for $m = 2$, the single angle $\phi_1 = \pi/2$ yields at most 1-HSE;
    \item for $m = 3$, choosing $\phi_1 = \pi/2$ for the odd degrees and $\phi_2$ at a zero of $P_2$ yields at most 3-HSE;
    \item for $m = 4$, choosing additionally $\phi_3$ at a zero of $P_4$ yields at most 5-HSE;
    \item and so on, each extra tone contributing one new angle and eliminating one further even degree.
\end{itemize}
Therefore,
\begin{equation}
    k_{\max} = 2m - 3 .
    \label{eq:S-kmax}
\end{equation}
The construction saturating this bound only requires locating one zero of each even-degree Legendre polynomial up to $P_{2(m-2)}$; an individual zero of $P_\ell$ can be evaluated to machine precision in $O(1)$ operations by iteration-free asymptotic expansions~\cite{SM-Bogaert2014}, so the $m-2$ angles cost $O(m) = O(k)$ in total, the cost quoted in the main text.

\subsection{Invariance under permutations of the inter-axis angles}
\label{sec:S2C}

The conditions~\eqref{eq:S-legendre-conditions} depend on the drive geometry only through the product $\prod_{j=1}^{m-1} P_\ell(\cos\phi_j)$, which is symmetric under any permutation of the angles $\{\phi_1, \ldots, \phi_{m-1}\}$.
Two $m$-tone drives whose consecutive-axis angles realize the same multiset of values therefore satisfy exactly the same set of conditions and achieve the same HSE order, regardless of the order in which the angles appear along the chain of axes.
Only the rank-one operator $|\ell, 0\rangle_{\boldsymbol{n}_m}~{}_{\boldsymbol{n}_1}\langle \ell, 0|$ in Eq.~\eqref{eq:S-nu-final} changes under such a rearrangement, and it never vanishes, so it plays no role in the conditions.
Hence any permutation of the angles $\phi_j$ leaves the HSE order of the drive unchanged, as used in the main text.
Quantities beyond the HSE order can, by contrast, depend on the arrangement: the fingerprint landscape of Sec.~\ref{sec:S3}, for instance, is organized around the first rotation axis.

\subsection{Additional examples of $m$-tone drives}
\label{sec:S2D}

\emph{A 2-tone drive realizing 1-HSE.}---The minimal example is the two-tone protocol of our previous work~\cite{SM-Liu2026},
\begin{equation}
    U(t) = e^{-i\omega_2 t  \sigma_z}  e^{-i t  \sigma_x} ,
    \label{eq:S-2tone}
\end{equation}
whose single inter-axis angle equals $\pi/2$.
By Eq.~\eqref{eq:S-legendre-conditions} it satisfies all odd-degree conditions and no even-degree one, realizing 1-HSE but not 2-HSE, in accordance with $k_{\max} = 2\times 2 - 3 = 1$.

\emph{The 3-tone drive of the main text.}---The experimental drive is the $m = 3$ instance with axes
\begin{equation}
    \boldsymbol{n}_1 = (1, 0, 0), \qquad
    \boldsymbol{n}_2 = \left(\frac{1}{\sqrt{3}},  \sqrt{\frac{2}{3}},  0\right), \qquad
    \boldsymbol{n}_3 = \left(-\sqrt{\frac{2}{3}},  \frac{1}{\sqrt{3}},  0\right),
    \label{eq:S-3tone-axes}
\end{equation}
and frequencies $\omega_1 = 1$, $\omega_2 = (1+\sqrt{2})/2$, $\omega_3 = (1+\sqrt{3})/2$, so that $\cos\phi_1 = \boldsymbol{n}_1\cdot\boldsymbol{n}_2 = 1/\sqrt{3}$ and $\phi_2 = \angle(\boldsymbol{n}_2, \boldsymbol{n}_3) = \pi/2$.
The right angle satisfies every odd-degree condition and $P_2(1/\sqrt{3}) = 0$ covers degree two, so the conditions~\eqref{eq:S-legendre-conditions} hold for $\ell = 1, 2, 3$ and the drive realizes 3-HSE, saturating $k_{\max} = 2\times 3 - 3 = 3$.
At degree four the product survives, $P_4(1/\sqrt{3})  P_4(0) = \big({-\tfrac{7}{18}}\big) \times \tfrac{3}{8} = -\tfrac{7}{48}$, and this surviving factor sets the amplitude of the fourth-order fingerprint derived in Sec.~\ref{sec:S3} and displayed in Fig.~1(d) of the main text.

\begin{figure}[!b]
\centering
\includegraphics[width=0.925\textwidth]{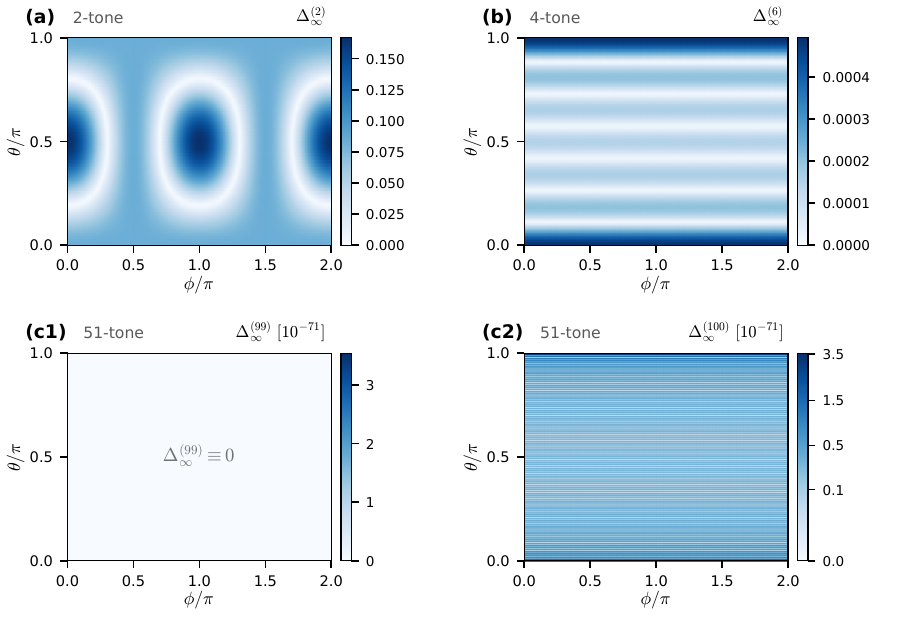}
\caption{
\textbf{Fingerprint landscapes of the additional drive examples.}
(a)~$\Delta_\infty^{(2)}$ of the 2-tone drive of Eq.~\eqref{eq:S-2tone}, which realizes 1-HSE, as a function of the initial state parameterized by the polar and azimuthal angles $(\theta, \phi)$.
(b)~$\Delta_\infty^{(6)}$ of the 4-tone drive of Eq.~\eqref{eq:S-4tone-drive}, which realizes 5-HSE.
(c1),(c2)~$\Delta_\infty^{(99)}$ and $\Delta_\infty^{(100)}$ of the 51-tone drive realizing 99-HSE, in units of $10^{-71}$: the former vanishes identically, while the latter displays the one hundred nodal rings of Eq.~\eqref{eq:S-51tone-A} on a nonlinear color scale that renders the interior oscillation visible.
The values in (a) and (b) are computed by numerical evaluation of the torus average; those in (c1) and (c2) follow the closed form of Eq.~\eqref{eq:S-51tone-A}.
The pattern of each landscape is organized around the first rotation axis of its drive, being concentric around the $\pm x$ directions in (a) and independent of $\phi$ in (b) and (c2), where the first axis is the $z$ axis.
}
\label{fig:S-extra-maps}
\end{figure}

\emph{A 4-tone drive realizing 5-HSE.}---For $m = 4$, we choose $\phi_1 = \pi/2$ for the odd degrees, $\phi_2$ at a zero of $P_2$, and $\phi_3$ at a zero of $P_4$.
An explicit set of axes is constructed as follows.
Take
\begin{equation}
    \boldsymbol{n}_1 = (0, 0, 1), \qquad \boldsymbol{n}_2 = (1, 0, 0),
    \label{eq:S-4tone-n12}
\end{equation}
so that $\phi_1 = \angle(\boldsymbol{n}_1, \boldsymbol{n}_2) = \pi/2$.
Next, choose $\boldsymbol{n}_3$ such that $\phi_2 = \angle(\boldsymbol{n}_2, \boldsymbol{n}_3)$ satisfies
\begin{equation}
    P_2(\cos\phi_2) = \frac{3\cos^2\phi_2 - 1}{2} = 0
    \quad\Longrightarrow\quad
    \boldsymbol{n}_2\cdot\boldsymbol{n}_3 = \cos\phi_2 = \frac{1}{\sqrt{3}} ,
    \label{eq:S-4tone-P2}
\end{equation}
for which it suffices to pick a vector in the $x$--$z$ plane with $x$ component $1/\sqrt{3}$,
\begin{equation}
    \boldsymbol{n}_3 = \left(\frac{1}{\sqrt{3}},  0,  \sqrt{\frac{2}{3}}\right).
    \label{eq:S-4tone-n3}
\end{equation}
Finally, choose $\boldsymbol{n}_4$ such that $\phi_3 = \angle(\boldsymbol{n}_3, \boldsymbol{n}_4)$ satisfies
\begin{equation}
    P_4(\cos\phi_3) = \frac{35\cos^4\phi_3 - 30\cos^2\phi_3 + 3}{8} = 0
    \quad\Longrightarrow\quad
    \cos\phi_3 = \pm\sqrt{\frac{15 \pm 2\sqrt{30}}{35}} .
    \label{eq:S-4tone-P4}
\end{equation}
Taking the root $\cos\phi_3 = \sqrt{(15 - 2\sqrt{30})/35}$ and writing $\boldsymbol{n}_4 = \cos\phi_3  \boldsymbol{n}_3 + s (0, 1, 0)$ with $s$ fixed by normalization,
\begin{equation}
    \boldsymbol{n}_4 = \left(\sqrt{\frac{15-2\sqrt{30}}{105}},  \sqrt{\frac{20+2\sqrt{30}}{35}},  \sqrt{\frac{2 (15-2\sqrt{30})}{105}}\right).
    \label{eq:S-4tone-n4}
\end{equation}
All rotation axes are then fixed, and the 4-tone protocol
\begin{equation}
    U(t) = e^{-i\omega_4 t  \boldsymbol{n}_4\cdot\boldsymbol{\sigma}}  e^{-i\omega_3 t  \boldsymbol{n}_3\cdot\boldsymbol{\sigma}}  e^{-i\omega_2 t  \sigma_x}  e^{-i\omega_1 t  \sigma_z} ,
    \label{eq:S-4tone-drive}
\end{equation}
with rationally independent frequencies, for instance $\{\omega_j\} = \{\sqrt{2}, \sqrt{3}, \sqrt{5}, \sqrt{7}\}$, achieves 5-HSE.
Numerical simulations confirm that this drive realizes 5-HSE but not 6-HSE.

\emph{A 51-tone drive realizing 99-HSE.}---The recipe extends to arbitrarily high order.
For $m = 51$ the bound~\eqref{eq:S-kmax} permits $k_{\max} = 99$: we keep $\phi_1 = \pi/2$ for the odd degrees and let the remaining angles $\phi_2, \ldots, \phi_{50}$ take one zero of each even-degree polynomial $P_2, P_4, \ldots, P_{98}$, the largest zero for definiteness.
A chain of axes realizing these angles is built exactly as in Eqs.~\eqref{eq:S-4tone-n12}--\eqref{eq:S-4tone-n4}, starting from $\boldsymbol{n}_1 = (0, 0, 1)$, and the $51$ rationally independent frequencies can be taken as the square roots of the first $51$ primes, $\{\omega_j\} = \{\sqrt{2}, \sqrt{3}, \sqrt{5}, \ldots, \sqrt{233}\}$.
Every condition with $\ell \le 99$ then holds, so the drive realizes 99-HSE and $\Delta_\infty^{(99)}$ vanishes identically over the Bloch sphere, whereas the $\ell = 100$ product survives at the value $\prod_{j=1}^{50} P_{100}(\cos\phi_j) \approx 5.1\times 10^{-42}\neq 0$.
The drive thus misses 100-HSE by a margin that is exactly computable yet very small; the resulting hundredth-order landscape is derived in Sec.~\ref{sec:S3} [Eq.~\eqref{eq:S-51tone-A}].

Figure~\ref{fig:S-extra-maps} collects the trace-distance landscapes of these drives: $\Delta_\infty^{(2)}$ of the 2-tone drive and $\Delta_\infty^{(6)}$ of the 4-tone drive are computed by numerical evaluation of the torus-average integral~\eqref{eq:S-torus-average}, a route independent of the analytic closed forms of Sec.~\ref{sec:S3} with which they agree, whereas $\Delta_\infty^{(99)}$ and $\Delta_\infty^{(100)}$ of the 51-tone drive, beyond the reach of numerical integration, are evaluated from the closed form of Sec.~\ref{sec:S3} [Eq.~\eqref{eq:S-51tone-A}]; the corresponding landscape of the 3-tone drive, $\Delta_\infty^{(4)}$, is displayed in Fig.~1(d) of the main text.
At the first broken order every landscape follows the $|P_{k+1}(\cos\gamma)|$ structure derived in Sec.~\ref{sec:S3}, organized around the first rotation axis of the drive: the first axes of the 2-tone and 3-tone drives are the $x$ axis, producing concentric patterns around $(\theta, \phi) = (\pi/2, 0)$ and $(\pi/2, \pi)$, while the first axes of the 4-tone and 51-tone drives coincide with the $z$ axis, producing bands independent of $\phi$.

\section{The $(k{+}1)$th-Order Deviation of $k$-HSE Dynamics}
\label{sec:S3}

The main text reports that the fourth-order deviation of the experimental 3-tone drive follows the closed form $\Delta_\infty^{(4)} = |P_4(\cos\gamma)|/60$ for every initial state, with $\gamma$ the angle between the Bloch vector and the first rotation axis.
This section derives the closed form of the $(k{+}1)$th-order deviation for a general $m$-tone drive realizing $k$-HSE, evaluates it for the drives of this work, and analyzes the structure of the resulting fingerprint landscape.

\subsection{General closed form of the $(k{+}1)$th-order deviation}
\label{sec:S3A}

Consider an $m$-tone drive that satisfies the conditions~\eqref{eq:S-legendre-conditions} for all $\ell \le k$ and fails at $\ell = k{+}1$, so that $k$-HSE holds but $(k{+}1)$-HSE does not; the saturating construction of Sec.~\ref{sec:S2B} realizes $k = 2m{-}3$.
For the drives of this work the first unmet degree $k{+}1$ is even, because a right angle among the inter-axis angles enforces every odd-degree condition; the derivation below, however, uses only the defining property that the conditions hold up to $\ell = k$ and fail at $k{+}1$.

By the definition~\eqref{eq:S-nu-def} of the temporal measure, the infinite-time moment at $k{+}1$ copies is the twirl of the $(k{+}1)$-copy initial state,
\begin{equation}
    \rho_\infty^{(k+1)} = \lim_{T\to\infty} \rho_T^{(k+1)} = \mathcal{T}^{(k+1)}_{\nu}\big(\rho_0^{\otimes(k+1)}\big) ,
    \label{eq:S-cf-setup}
\end{equation}
and, applying Secs.~\ref{sec:S1B} and \ref{sec:S1C} with the copy number now $k{+}1$ in place of $k$, the input $\rho_0^{\otimes(k+1)}$, the two twirls, and their outputs can all be restricted to the symmetric subspace $V_{j'}$ with $j' = (k{+}1)/2$, of dimension $2j' + 1 = k + 2$, on which the Haar moment is $\rho_{\mathrm{Haar}}^{(k+1)} = I_{V_{j'}}/(k{+}2)$ with $I_{V_{j'}}$ the identity on $V_{j'}$, because the Haar Fourier components~\eqref{eq:S-haar-fourier} annihilate every sector except $\ell = 0$.
The operator space accordingly decomposes as $\mathrm{Hom}(V_{j'}, V_{j'}) \cong \bigoplus_{\ell=0}^{k+1} V_\ell$ by Eq.~\eqref{eq:S-CG}, so the spherical tensor basis of Sec.~\ref{sec:S1D} now carries the ranks $\ell = 0, 1, \ldots, k{+}1$.
Expanding the input as $\rho_0^{\otimes(k+1)} = \sum_{\ell=0}^{k+1}\sum_m x_{\ell m} T^{(\ell)}_m$, with $x_{\ell m} = \mathrm{tr}\big[T^{(\ell)\dagger}_m \rho_0^{\otimes(k+1)}\big]$ as in Sec.~\ref{sec:S1D}, applying the transformation law~\eqref{eq:S-tensor-transform} to each term, and noting that the $U$ dependence resides entirely in the scalar matrix elements $D^{(\ell)}_{m'm}(U)$,
\begin{equation}
\begin{aligned}
    \rho_\infty^{(k+1)}
    &= \int_{U\sim\nu} D^{(j')}(U)  \rho_0^{\otimes(k+1)}  D^{(j')}(U)^\dagger \mathrm{d}\nu
    = \sum_{\ell=0}^{k+1}\sum_{m=-\ell}^{\ell} x_{\ell m} \int_{U\sim\nu} D^{(j')}(U)  T^{(\ell)}_m  D^{(j')}(U)^\dagger \mathrm{d}\nu \\
    &= \sum_{\ell=0}^{k+1}\sum_{m, m'} x_{\ell m} \left(\int_{U\sim\nu} D^{(\ell)}_{m'm}(U) \mathrm{d}\nu\right) T^{(\ell)}_{m'}
    = \sum_{\ell=0}^{k+1} \sum_{m, m'} \big[\hat{\nu}(\ell)\big]_{m'm}  x_{\ell m}  T^{(\ell)}_{m'} ,
\end{aligned}
    \label{eq:S-cf-sectorwise}
\end{equation}
so the twirl acts on each sector separately, through the same Fourier components $\hat{\nu}(\ell)$ that entered the $k$-HSE conditions.
The sectors fall into three groups.
The $\ell = 0$ term carries $T^{(0)}_0 = I_{V_{j'}}/\sqrt{k+2}$ with coefficient $x_{00} = \mathrm{tr}[T^{(0)\dagger}_0 \rho_0^{\otimes(k+1)}] = 1/\sqrt{k+2}$, and $\hat{\nu}(0) = 1$ leaves it untouched, reproducing exactly the Haar moment.
The sectors $\ell = 1, \ldots, k$ are annihilated, because the drive satisfies $k$-HSE and hence $\hat{\nu}(\ell) = 0$ there.
The only survivor beyond the Haar part is the top rank $\ell = k{+}1 = 2j'$, where Eq.~\eqref{eq:S-nu-final} gives the rank-one Fourier component
\begin{equation}
    \hat{\nu}(2j') = \Bigg[\prod_{i=1}^{m-1} P_{2j'}(\cos\phi_i)\Bigg] |2j', 0\rangle_{\boldsymbol{n}_m}~{}_{\boldsymbol{n}_1}\langle 2j', 0| .
    \label{eq:S-cf-nu-top}
\end{equation}
To read off its action we introduce frame-adapted tensors: for a unit vector $\boldsymbol{n}$ with $R_{\boldsymbol{n}}  \sigma_z  R_{\boldsymbol{n}}^\dagger = \boldsymbol{n}\cdot\boldsymbol{\sigma}$, let $T^{(\ell)}_m[\boldsymbol{n}] = D^{(j')}(R_{\boldsymbol{n}})  T^{(\ell)}_m  D^{(j')}(R_{\boldsymbol{n}})^\dagger$ denote the spherical tensors whose weight is measured along $\boldsymbol{n}$, and let $x_{\ell m}[\boldsymbol{n}]$ be the expansion coefficients of $\rho_0^{\otimes(k+1)}$ in this basis.
The rotated tensors $\{T^{(\ell)}_m[\boldsymbol{n}]\}$ form again an orthonormal basis, because unitary conjugation preserves the Hilbert--Schmidt inner product, so the frame-$\boldsymbol{n}$ coefficients follow by projection,
\begin{equation}
\begin{aligned}
    x_{\ell m'}[\boldsymbol{n}]
    &= \mathrm{tr}\big[ T^{(\ell)}_{m'}[\boldsymbol{n}]^\dagger  \rho_0^{\otimes(k+1)} \big]
    = \sum_{m=-\ell}^{\ell} \big[D^{(\ell)}_{m m'}(R_{\boldsymbol{n}})\big]^*  \mathrm{tr}\big[ T^{(\ell)\dagger}_{m}  \rho_0^{\otimes(k+1)} \big] \\
    &= \sum_{m=-\ell}^{\ell} \big[D^{(\ell)}(R_{\boldsymbol{n}})^\dagger\big]_{m'm}  x_{\ell m}
    = \sum_{m=-\ell}^{\ell} {}_{\boldsymbol{n}}\langle \ell, m'|\ell, m\rangle  x_{\ell m} ,
\end{aligned}
    \label{eq:S-cf-frame-coeff}
\end{equation}
where the second equality expands $T^{(\ell)}_{m'}[\boldsymbol{n}] = D^{(j')}(R_{\boldsymbol{n}})  T^{(\ell)}_{m'}  D^{(j')}(R_{\boldsymbol{n}})^\dagger = \sum_m D^{(\ell)}_{m m'}(R_{\boldsymbol{n}})  T^{(\ell)}_m$ by the transformation law~\eqref{eq:S-tensor-transform} evaluated at $U = R_{\boldsymbol{n}}$, with the scalar matrix elements $D^{(\ell)}_{m m'}(R_{\boldsymbol{n}})$ passing out of the trace, the third recognizes the traces as the $z$-frame coefficients $x_{\ell m}$, and the last writes the Wigner matrix element as the frame overlap $\big[D^{(\ell)}(R_{\boldsymbol{n}})^\dagger\big]_{m'm} = \langle \ell, m'| D^{(\ell)}(R_{\boldsymbol{n}})^\dagger |\ell, m\rangle = {}_{\boldsymbol{n}}\langle \ell, m'|\ell, m\rangle$; the coefficients $x_{\ell m'}[\boldsymbol{n}]$ at fixed $\ell$ thus transform as the components of a spin-$\ell$ ket.
In the basis $\{|2j', m\rangle\}$ of weights along $z$, with states carrying no frame subscript always referring to the $z$ frame, the matrix elements of Eq.~\eqref{eq:S-cf-nu-top} factorize into a product of two overlaps,
\begin{equation}
    \big[\hat{\nu}(2j')\big]_{m'm}
    = \langle 2j', m'|  \hat{\nu}(2j')  |2j', m\rangle
    = \Bigg[\prod_{i=1}^{m-1} P_{2j'}(\cos\phi_i)\Bigg] \langle 2j', m'|2j', 0\rangle_{\boldsymbol{n}_m}  {}_{\boldsymbol{n}_1}\langle 2j', 0|2j', m\rangle .
    \label{eq:S-cf-nu-matelem}
\end{equation}
Subtracting the Haar part leaves the single sector $\ell = 2j'$ of Eq.~\eqref{eq:S-cf-sectorwise}, and inserting Eq.~\eqref{eq:S-cf-nu-matelem} splits the double sum, collapsing the deviation to a single term,
\begin{equation}
\begin{aligned}
    \rho_\infty^{(k+1)} - \rho_{\mathrm{Haar}}^{(k+1)}
    &= \sum_{m, m'} \big[\hat{\nu}(2j')\big]_{m'm}  x_{2j' m}  T^{(2j')}_{m'} \\
    &= \Bigg[\prod_{i=1}^{m-1} P_{2j'}(\cos\phi_i)\Bigg]
    \Bigg[\sum_{m} {}_{\boldsymbol{n}_1}\langle 2j', 0|2j', m\rangle  x_{2j' m}\Bigg]
    \Bigg[\sum_{m'} \langle 2j', m'|2j', 0\rangle_{\boldsymbol{n}_m}  T^{(2j')}_{m'}\Bigg] \\
    &\equiv \Bigg[\prod_{i=1}^{m-1} P_{2j'}(\cos\phi_i)\Bigg]  x_{2j',0}[\boldsymbol{n}_1]  T^{(2j')}_0[\boldsymbol{n}_m] .
\end{aligned}
    \label{eq:S-cf-deviation}
\end{equation}
The last line identifies the two brackets: the first is the projection~\eqref{eq:S-cf-frame-coeff} evaluated at $\ell = 2j'$, $m' = 0$, $\boldsymbol{n} = \boldsymbol{n}_1$, the zero-weight coefficient of the input in the $\boldsymbol{n}_1$ frame, and the second is the transformation law~\eqref{eq:S-tensor-transform} applied to $T^{(2j')}_0$ at the rotation $R_{\boldsymbol{n}_m}$, that is, the zero-weight tensor along $\boldsymbol{n}_m$.

The initial state enters Eq.~\eqref{eq:S-cf-deviation} only through the scalar $x_{2j',0}[\boldsymbol{n}_1]$.
Writing $|\psi(0)\rangle = R_{\boldsymbol{r}} |0\rangle$ up to an irrelevant global phase, where $\boldsymbol{r}$ is the unit Bloch vector of the pure initial state and $R_{\boldsymbol{r}}$ is the rotation $R_{\boldsymbol{n}}$ introduced above, evaluated at $\boldsymbol{n} = \boldsymbol{r}$, the $(k{+}1)$-copy state restricted to $V_{j'}$ is a rotated highest-weight state,
\begin{equation}
    \rho_0^{\otimes(k+1)}\big|_{V_{j'}} = D^{(j')}(R_{\boldsymbol{r}})  |j', j'\rangle\langle j', j'|  D^{(j')}(R_{\boldsymbol{r}})^\dagger ,
    \label{eq:S-cf-coherent}
\end{equation}
because $|0\rangle^{\otimes(k+1)}$ is the eigenstate of $J_z \equiv J_z^{(j')}$, the $z$ angular momentum on $V_{j'}$ in the notation of Eq.~\eqref{eq:S-phase-action}, with the maximal eigenvalue, namely the highest-weight state $|j', j'\rangle$.
The projector $|j', j'\rangle\langle j', j'|$ commutes with $J_z$, hence with the rotation operator $e^{-i2\theta J_z} = D^{(j')}\big(e^{-i\theta\sigma_z}\big)$ at every angle $\theta$, which is to say that it is invariant under all rotations about the $z$ axis,
\begin{equation}
    e^{-i2\theta J_z}  |j', j'\rangle\langle j', j'|  e^{+i2\theta J_z} = |j', j'\rangle\langle j', j'| .
    \label{eq:S-cf-z-invariance}
\end{equation}
The basis tensors respond to the same rotations with a pure phase: evaluating the transformation law~\eqref{eq:S-tensor-transform} at $U = e^{-i\theta\sigma_z}$, whose spin-$\ell$ matrix elements are diagonal by Eq.~\eqref{eq:S-phase-action}, $D^{(\ell)}_{m'm}\big(e^{-i\theta\sigma_z}\big) = e^{-i2m\theta}  \delta_{m'm}$, collapses the sum over $m'$ to the single term
\begin{equation}
    e^{-i2\theta J_z}  T^{(\ell)}_m  e^{+i2\theta J_z} = e^{-i2m\theta}  T^{(\ell)}_m .
    \label{eq:S-cf-weight-phase}
\end{equation}
Now expand the projector in the tensor basis as $|j', j'\rangle\langle j', j'| = \sum_{\ell=0}^{2j'}\sum_{m=-\ell}^{\ell} a_{\ell m}  T^{(\ell)}_m$, substitute the expansion into the left-hand side of Eq.~\eqref{eq:S-cf-z-invariance}, and apply Eq.~\eqref{eq:S-cf-weight-phase} term by term:
\begin{equation}
    \sum_{\ell=0}^{2j'}\sum_{m=-\ell}^{\ell} a_{\ell m}  e^{-i2m\theta}  T^{(\ell)}_m
    = \sum_{\ell=0}^{2j'}\sum_{m=-\ell}^{\ell} a_{\ell m}  T^{(\ell)}_m .
    \label{eq:S-cf-coeff-match}
\end{equation}
Since the operators $T^{(\ell)}_m$ are linearly independent, the coefficients on the two sides must agree one by one, $a_{\ell m}  e^{-i2m\theta} = a_{\ell m}$ for every $\theta$; for $m \neq 0$ the choice $\theta = \pi/(2m)$ gives $a_{\ell m} = -a_{\ell m}$ and hence $a_{\ell m} = 0$, while the coefficients at $m = 0$ are unrestricted.
The expansion in the $z$ frame therefore contains zero-weight tensors only,
\begin{equation}
    |j', j'\rangle\langle j', j'| = \sum_{\ell=0}^{2j'} a_\ell  T^{(\ell)}_0 ,
    \label{eq:S-cf-axial}
\end{equation}
where the coefficients follow from the Hilbert--Schmidt projection onto the orthonormal basis, with the trace against the rank-one projector collapsing to an expectation value,
\begin{equation}
    a_\ell = \mathrm{tr}\big[ T^{(\ell)\dagger}_0  |j', j'\rangle\langle j', j'| \big]
    = \langle j', j'|  T^{(\ell)\dagger}_0  |j', j'\rangle .
    \label{eq:S-cf-axial-coeff}
\end{equation}
Substituting Eq.~\eqref{eq:S-cf-axial} into Eq.~\eqref{eq:S-cf-coherent}, the rotation acts on each term of the expansion and turns every $T^{(\ell)}_0$ into its $\boldsymbol{r}$-frame counterpart,
\begin{equation}
    \rho_0^{\otimes(k+1)}\big|_{V_{j'}}
    = D^{(j')}(R_{\boldsymbol{r}}) \Bigg[\sum_{\ell=0}^{2j'} a_\ell  T^{(\ell)}_0\Bigg] D^{(j')}(R_{\boldsymbol{r}})^\dagger
    = \sum_{\ell=0}^{2j'} a_\ell  T^{(\ell)}_0[\boldsymbol{r}] .
    \label{eq:S-cf-input-expanded}
\end{equation}
Tensors $T^{(\ell)}_m[\boldsymbol{n}]$ of different rank $\ell$ are orthogonal in the Hilbert--Schmidt inner product regardless of their frames $\boldsymbol{n}$, because a rotation mixes each rank only within itself by the transformation law~\eqref{eq:S-tensor-transform}, so projecting onto the top-rank zero-weight tensor of the $\boldsymbol{n}_1$ frame picks up a single term,
\begin{equation}
    x_{2j',0}[\boldsymbol{n}_1]
    = \mathrm{tr}\Big[ T^{(2j')}_0[\boldsymbol{n}_1]^\dagger  \rho_0^{\otimes(k+1)}\Big]
    = a_{2j'}  \mathrm{tr}\Big[ T^{(2j')}_0[\boldsymbol{n}_1]^\dagger  T^{(2j')}_0[\boldsymbol{r}]\Big]
    = a_{2j'}  d^{2j'}_{00}(\gamma)
    = a_{2j'}  P_{2j'}(\cos\gamma) ,
    \qquad
    \gamma = \angle(\boldsymbol{r}, \boldsymbol{n}_1) ,
    \label{eq:S-cf-scalar}
\end{equation}
where the third equality expands $T^{(2j')}_0[\boldsymbol{r}]$ in the $\boldsymbol{n}_1$ frame and keeps the $m = 0$ term of the trace, the same Wigner-$d$ evaluation as in Eq.~\eqref{eq:S-legendre-overlap}.
The initial state thus enters the deviation through exactly the same Legendre mechanism as the drive angles, with the polar angle $\gamma$ playing the role of an inter-axis angle between the Bloch vector and the first rotation axis; and since $\gamma$ is the only state dependence, the landscape is axially symmetric about $\boldsymbol{n}_1$: the temporal ensemble retains the latitude of the Bloch vector relative to $\boldsymbol{n}_1$ and forgets the longitude around it, as stated in the main text.
Since $\rho_T^{(k+1)} \to \rho_\infty^{(k+1)}$ by Eq.~\eqref{eq:S-cf-setup} and the trace norm is continuous on the finite-dimensional space of operators, the limit of the trace distances is the trace distance of the limit, $\Delta_\infty^{(k+1)} = \tfrac{1}{2}\big\|\rho_\infty^{(k+1)} - \rho_{\mathrm{Haar}}^{(k+1)}\big\|_1$, and taking one half of the trace norm of Eq.~\eqref{eq:S-cf-deviation} with the unitary invariance $\|T^{(2j')}_0[\boldsymbol{n}_m]\|_1 = \|T^{(2j')}_0\|_1$,
\begin{equation}
    \Delta_\infty^{(k+1)}
    = \frac{1}{2}  \Bigg|\prod_{i=1}^{m-1} P_{2j'}(\cos\phi_i)\Bigg|  |a_{2j'}|  \big\|T^{(2j')}_0\big\|_1  \big|P_{2j'}(\cos\gamma)\big| .
    \label{eq:S-cf-assembled}
\end{equation}

The two remaining constants of Eq.~\eqref{eq:S-cf-assembled}, $|a_{2j'}|$ and $\big\|T^{(2j')}_0\big\|_1$, follow from the explicit form of $T^{(2j')}_0$.
Each zero-weight tensor $T^{(\ell)}_0$ with $\ell = 0, \ldots, 2j'$---an operator on $V_{j'}$ that transforms with rank $\ell$ under Eq.~\eqref{eq:S-tensor-transform}---is invariant under every rotation about the $z$ axis by Eq.~\eqref{eq:S-cf-weight-phase} at $m = 0$, hence commutes with $J_z$ and is diagonal in the basis $\{|j', m\rangle\}$.
The diagonal operators on $V_{j'}$ form a $(2j'{+}1)$-dimensional space with two natural bases: the powers $J_z^{ p}$ with $p = 0, \ldots, 2j'$, and the zero-weight tensors $T^{(\ell)}_0$, one from each sector.
Since $J_z \propto T^{(1)}_0$ and a product of $p$ rank-one tensors decomposes into ranks at most $p$, the span of $\{J_z^{ 0}, \ldots, J_z^{ p}\}$ lies in that of $\{T^{(0)}_0, \ldots, T^{(p)}_0\}$, and matching dimensions makes the two spans equal for every $p$.
As a member of the orthonormal basis, the top tensor $T^{(2j')}_0$ is orthogonal to all the lower zero-weight tensors, $\mathrm{tr}\big[T^{(2j')\dagger}_0  T^{(\ell)}_0\big] = 0$ for $\ell = 0, \ldots, 2j'{-}1$---a condition that is hard to use directly, because the intermediate tensors $T^{(1)}_0, \ldots, T^{(2j'-1)}_0$ have not been constructed explicitly.
The equality of the two spans at $p = 2j'{-}1$ removes this obstacle: orthogonality to the lower tensors is equivalent to orthogonality to the explicitly known powers of $J_z$,
\begin{equation}
    \mathrm{tr}\big[ T^{(2j')}_0  J_z^{ p} \big] = 0 ,
    \qquad p = 0, 1, \ldots, 2j'{-}1 .
    \label{eq:S-cf-power-orth}
\end{equation}
These are $2j'$ independent linear constraints on the $(2j'{+}1)$-dimensional space of diagonal operators, so together with the normalization they single out $T^{(2j')}_0$ uniquely, up to sign.
It therefore suffices to exhibit one normalized diagonal operator satisfying Eq.~\eqref{eq:S-cf-power-orth}, and the alternating binomial profile does,
\begin{equation}
    T^{(2j')}_0 = \frac{1}{\sqrt{\binom{4j'}{2j'}}} \sum_{m=-j'}^{j'} (-1)^{j'-m} \binom{2j'}{j'+m}  |j', m\rangle\langle j', m| ,
    \label{eq:S-cf-T-top}
\end{equation}
as we now verify by reducing the orthogonality to a finite-difference identity.
Both operators in Eq.~\eqref{eq:S-cf-power-orth} are diagonal, with entries read off Eq.~\eqref{eq:S-cf-T-top} and $m^p$ respectively, so the trace is a single sum over the diagonal,
\begin{equation}
    \mathrm{tr}\big[ T^{(2j')}_0  J_z^{ p} \big]
    = \frac{1}{\sqrt{\binom{4j'}{2j'}}} \sum_{m=-j'}^{j'} (-1)^{j'-m} \binom{2j'}{j'+m}  m^p
    = \frac{1}{\sqrt{\binom{4j'}{2j'}}} \sum_{i=0}^{2j'} (-1)^{i} \binom{2j'}{i}  (j' - i)^p ,
    \label{eq:S-cf-orth-sum}
\end{equation}
where the second equality substitutes $i = j' - m$.
The sum vanishes by an algebraic identity of alternating binomial sums: for every real $x$ and all integers $0 \le m < n$,
\begin{equation}
    \sum_{i=0}^{n} (-1)^i \binom{n}{i}  (x - i)^m = 0 .
    \label{eq:S-cf-finite-diff}
\end{equation}
The sum in Eq.~\eqref{eq:S-cf-orth-sum} is precisely the case $n = 2j'$, $x = j'$, and $m = p \le 2j' - 1 < n$, so the trace vanishes for every $p$ and Eq.~\eqref{eq:S-cf-power-orth} is verified.
Finally, the prefactor of Eq.~\eqref{eq:S-cf-T-top} ensures the normalization: the operator is diagonal, so its squared Hilbert--Schmidt norm is the sum of its squared diagonal entries,
\begin{equation}
    \mathrm{tr}\big[ T^{(2j')\dagger}_0  T^{(2j')}_0 \big]
    = \frac{1}{\binom{4j'}{2j'}} \sum_{m=-j'}^{j'} \binom{2j'}{j'+m}^2
    = 1 ,
    \label{eq:S-cf-normalization}
\end{equation}
where the second equality is the Vandermonde identity $\sum_{m} \binom{2j'}{j'+m}^2 = \binom{4j'}{2j'}$.

Reading off the two constants from Eq.~\eqref{eq:S-cf-T-top},
\begin{equation}
    a_{2j'} = \langle j', j'|  T^{(2j')}_0  |j', j'\rangle = \frac{1}{\sqrt{\binom{4j'}{2j'}}} ,
    \qquad
    \big\|T^{(2j')}_0\big\|_1 = \frac{1}{\sqrt{\binom{4j'}{2j'}}} \sum_{m=-j'}^{j'} \binom{2j'}{j'+m} = \frac{4^{j'}}{\sqrt{\binom{4j'}{2j'}}} ,
    \label{eq:S-cf-constants}
\end{equation}
and substituting them into Eq.~\eqref{eq:S-cf-assembled} yields the general closed form
\begin{equation}
    \Delta_\infty^{(2j')}
    = \frac{1}{2}  \frac{4^{j'}}{\binom{4j'}{2j'}}  \Bigg|\prod_{i=1}^{m-1} P_{2j'}(\cos\phi_i)\Bigg|  \big|P_{2j'}(\cos\gamma)\big| ,
    \label{eq:S-cf-general}
\end{equation}
or, written entirely in terms of the HSE order $k = 2j' - 1$,
\begin{equation}
    \Delta_\infty^{(k+1)}
    = \frac{1}{2}  \frac{2^{k+1}}{\binom{2k+2}{k+1}}  \Bigg|\prod_{i=1}^{m-1} P_{k+1}(\cos\phi_i)\Bigg|  \big|P_{k+1}(\cos\gamma)\big| ,
    \label{eq:S-cf-general-k}
\end{equation}
with $\gamma$ measured from the first rotation axis of the drive.
The structure is universal: the polynomial of the first unmet condition shapes both sides of the deviation, with the drive angles entering through the amplitude factor $\prod_i P_{k+1}(\cos\phi_i)$ and the initial state contributing the fingerprint $P_{k+1}(\cos\gamma)$.

\subsection{Evaluation for the drives of this work}
\label{sec:S3B}

For the experimental 3-tone drive, $k = 3$ and the prefactor of Eq.~\eqref{eq:S-cf-general-k} is $\frac{1}{2}\cdot\frac{16}{70} = \frac{4}{35}$,
\begin{equation}
    \Delta_\infty^{(4)}
    = \frac{4}{35}  \big|P_4(\cos\phi_1)  P_4(\cos\phi_2)\big|  \big|P_4(\cos\gamma)\big| ,
    \label{eq:S-cf-general-form}
\end{equation}
with the two drive factors
\begin{equation}
    P_4\Big(\frac{1}{\sqrt{3}}\Big) = \frac{1}{8}\Big(\frac{35}{9} - 10 + 3\Big) = -\frac{7}{18},
    \qquad
    P_4(0) = \frac{3}{8} ,
    \label{eq:S-cf-factors}
\end{equation}
giving the constant quoted in the main text,
\begin{equation}
    \Delta_\infty^{(4)}
    = \frac{4}{35}\cdot\frac{7}{18}\cdot\frac{3}{8}  \big|P_4(\cos\gamma)\big|
    = \frac{1}{60}  \big|P_4(\cos\gamma)\big| ,
    \label{eq:S-cf-closed-form}
\end{equation}
the landscape displayed in Fig.~1(d) of the main text.
An exact numerical evaluation of the torus average over random initial states reproduces Eq.~\eqref{eq:S-cf-closed-form}.

For the 2-tone drive of Sec.~\ref{sec:S2D}, $k = 1$ and the single angle $\pi/2$ give $\Delta_\infty^{(2)} = \frac{1}{2}\cdot\frac{4}{6}\cdot|P_2(0)| |P_2(\cos\gamma)| = |P_2(\cos\gamma)|/6$, whose maximum $1/6$ is the peak value of Fig.~\ref{fig:S-extra-maps}(a).
For the 4-tone drive, $k = 5$ and the three angles of Sec.~\ref{sec:S2D} give the peak value $\approx 4.94\times 10^{-4}$ of Fig.~\ref{fig:S-extra-maps}(b).
For the 51-tone drive, Eq.~\eqref{eq:S-cf-general-k} with $k = 99$ yields the exact landscape
\begin{equation}
    \Delta_\infty^{(100)} = A  \big|P_{100}(\cos\gamma)\big| ,
    \qquad
    A = \frac{1}{2}  \frac{2^{100}}{\binom{200}{100}} \prod_{i=1}^{50} \big|P_{100}(\cos\phi_i)\big| \approx 3.5\times 10^{-71} ,
    \label{eq:S-51tone-A}
\end{equation}
a pattern of one hundred nodal rings around $\boldsymbol{n}_1$ [Fig.~\ref{fig:S-extra-maps}(c2)], at an amplitude suppressed by the product of fifty Legendre factors.
The 2-tone and 4-tone maps of Fig.~\ref{fig:S-extra-maps}, computed by exact numerical evaluation of the torus average, agree with Eq.~\eqref{eq:S-cf-general-k}.
At $k = 99$ such a numerical evaluation is no longer feasible, so the 51-tone panels of Fig.~\ref{fig:S-extra-maps} display the closed form~\eqref{eq:S-51tone-A} itself, whose validity rests on the derivation of Sec.~\ref{sec:S3A} and its numerical confirmation at the lower orders.

\subsection{Structure of the fingerprint landscape}
\label{sec:S3C}

The radial profile of the fourth-order landscape is the oscillation of $|P_4(\cos\gamma)|/60$, and its features can be read off the polynomial.
Legendre polynomials obey $|P_\ell(x)| \le 1$ on $[-1, 1]$ with equality only at $x = \pm 1$, so the global maxima sit exactly at $\boldsymbol{r} = \pm\boldsymbol{n}_1$, the $\sigma_x$ eigenstates $|\pm\rangle = (|0\rangle \pm |1\rangle)/\sqrt{2}$, with the peak value $1/60$.
The mechanism behind the peaks is the one described in the main text: for these states the first rotation acts trivially, so the dynamics they experience is that of the 2-tone drive formed by the remaining two axes, the least ergodic member of the family.
The zeros of $P_4$, located at the values $\cos\gamma = \pm\sqrt{(15 \pm 2\sqrt{30})/35}$ that already appeared in Eq.~\eqref{eq:S-4tone-P4}, produce four nodal rings at $\gamma \approx 30.56^\circ$ and $70.12^\circ$ together with their mirrors at $109.88^\circ$ and $149.44^\circ$, on which $\Delta_\infty^{(4)}$ vanishes exactly.
Between the rings, the interior extrema of $P_4$ at $\cos\gamma = \pm\sqrt{3/7}$ produce secondary maximum rings at $\gamma \approx 49.11^\circ$ and its mirror $130.89^\circ$ with $|P_4| = 3/7$ and $\Delta_\infty^{(4)} = 1/140$, and the equatorial great circle $\gamma = \pi/2$, which contains $|0\rangle$, carries the local extremum $|P_4(0)| = 3/8$ with $\Delta_\infty^{(4)} = 1/160$.
The color maps of Fig.~1(d) of the main text use the polar and azimuthal angles $(\theta, \phi)$ of the Bloch sphere with polar axis $z$, while $\boldsymbol{n}_1$ points along $x$, so the two coordinate systems are related by $\cos\gamma = \sin\theta\cos\phi$: the peaks appear as bullseye centers at $(\theta, \phi) = (\pi/2, 0)$ and $(\pi/2, \pi)$, surrounded by the nodal and secondary rings.
The same reading applies to the other drives of Fig.~\ref{fig:S-extra-maps}, whose fingerprints $|P_{k+1}(\cos\gamma)|$ differ only in the degree of the polynomial.
The number of nodal rings equals the number of zeros of $P_{k+1}$, namely $k{+}1$, so the oscillation grows richer with the order: the 2-tone landscape at $k = 1$ shows the simplest structure, two nodal rings and a single secondary band between the two maxima, while the 4-tone and 51-tone landscapes carry six and one hundred nodal rings.
For these latter two drives the first axis points along $z$, so $\gamma$ coincides with $\theta$ and the rings appear as latitude bands independent of $\phi$, in contrast to the bullseye patterns of the 2-tone and 3-tone drives, whose first axes point along $x$.

\section{Experimental Methods}
\label{sec:S4}

\subsection{Diamond sample and NV center characterization}
\label{sec:S4A}

The experiment was performed on the same single nitrogen-vacancy (NV) center~\cite{SM-Doherty2013} in diamond and the same optically detected magnetic resonance setup as our previous work, and a complete characterization of the sample and the apparatus is given in the Supplemental Material of Ref.~\cite{SM-Liu2026}.
In brief, the diamond was grown by chemical vapor deposition on a [100]-oriented substrate, using isotopically purified methane with a $^{12}$C abundance of $99.9\%$, and the NV centers were created by $\mathrm{N}_2^+$ implantation at about $30~\mathrm{keV}$ with a dose of about $5\times 10^{8}~\mathrm{cm}^{-2}$, followed by annealing at $1000~^\circ\mathrm{C}$, resulting in near-surface centers at a depth of around $10~\mathrm{nm}$.
The qubit is encoded in the subspace $\{|0\rangle_{\mathrm{e}}, |{-}1\rangle_{\mathrm{e}}\}$ of the spin-1 electronic ground state, split by a static magnetic field.
The characterized coherence properties are $T_{1, |0\rangle_{\mathrm{e}}} = 3.7(1)~\mathrm{ms}$, $T_{1, |{-}1\rangle_{\mathrm{e}}} = 8(2)~\mathrm{ms}$, and $T_2^* = 68(3)~\mu\mathrm{s}$~\cite{SM-Liu2026}, all longer than the microwave sequences applied in this work, and the optically pumped spin polarization is $p_{\mathrm{e}} \approx 0.96$ in the encoding subspace.

\subsection{Pulse sequences and calibration}
\label{sec:S4B}

The three rotation axes of the drive lie in the equatorial plane of the Bloch sphere, so every factor of $U(t)$ can be realized by a single resonant microwave pulse.
The microwave field couples to the spin through the laboratory-frame Hamiltonian
\begin{equation}
    H = \omega_0 S_z + 2\Omega \cos(\omega_0 t' + \phi)  S_x ,
    \label{eq:S-lab-H}
\end{equation}
where $\omega_0$ is the resonance frequency of the encoding subspace, $t'$ the laboratory time, $\Omega$ the Rabi frequency, $\phi$ the pulse phase, and $S_{x,y,z} = \sigma_{x,y,z}/2$ the spin operators.
In the frame rotating at $\omega_0$, after the rotating-wave approximation, the Hamiltonian becomes
\begin{equation}
    H_{\mathrm{rot}} = \Omega  (\cos\phi  S_x + \sin\phi  S_y) ,
    \label{eq:S-rot-H}
\end{equation}
which generates a rotation about the equatorial axis of azimuth $\phi$ at the fixed rate $\Omega$: the pulse phase selects the rotation axis, and the pulse duration sets the rotation angle.
For the stroboscopic time $t$, the target unitary $U(t) = e^{-i\omega_3 t H_3}  e^{-i\omega_2 t H_2}  e^{-i\omega_1 t H_1}$ is implemented as three successive pulses, applied in reverse order of the factors, whose phases match the azimuths of $\boldsymbol{n}_1$, $\boldsymbol{n}_2$, and $\boldsymbol{n}_3$ and whose durations $\tau_j = 2 (\omega_j t \bmod 2\pi)/\Omega$ realize the rotation angles.
The Rabi frequency is fixed at $\Omega = 2\pi \times 0.5~\mathrm{MHz}$ throughout, and the drive frequencies are $\omega_1 = 1$, $\omega_2 = (1+\sqrt{2})/2$, and $\omega_3 = (1+\sqrt{3})/2$, as quoted in the main text.
A single experimental run consists of laser initialization, a preparation pulse rotating $|0\rangle$ to the target initial state $|\psi(0)\rangle$ where needed, the three drive pulses for the chosen $t$, one of the six tomography pulses described below, and optical readout.
The runs iterate over $t = 0, 1, \ldots, 1999$ for the two initial states $|0\rangle$ and $|\alpha\rangle$.

\subsection{Quantum state tomography and data processing}
\label{sec:S4C}

The spin state after each run is reconstructed from the state-dependent photoluminescence (PL) rate.
Writing $l_0$ and $l_1$ for the PL rates of $|0\rangle$ and $|1\rangle$, a density matrix
\begin{equation}
    \rho = \begin{pmatrix}
        p_0 & \alpha + i\beta \\
        \alpha - i\beta & p_1
    \end{pmatrix} ,
    \qquad p_0 + p_1 = 1 ,
    \label{eq:S-tomo-rho}
\end{equation}
produces the expected PL rate $E = p_0 l_0 + p_1 l_1$.
Six readout sequences are performed, applying to $\rho$ the identity ($E_1$), $R_X(\pi)$ ($E_2$), $R_X(\pi/2)$ ($E_3$), $R_{-X}(\pi/2)$ ($E_4$), $R_Y(\pi/2)$ ($E_5$), and $R_{-Y}(\pi/2)$ ($E_6$) before optical readout, where $R_{\boldsymbol{n}}(\theta) = e^{-i\theta  \boldsymbol{n}\cdot\boldsymbol{\sigma}/2}$ denotes the spin rotation about the axis $\boldsymbol{n}$ by the angle $\theta$; the rotations map the coherences onto the populations and yield
\begin{equation}
    \left\{
    \begin{aligned}
        & l_0  p_0 + l_1  p_1 = E_1 , \\
        & l_0  p_1 + l_1  p_0 = E_2 , \\
        & l_0  \frac{p_0 + p_1 - 2\beta}{2} + l_1  \frac{p_0 + p_1 + 2\beta}{2} = E_3 , \\
        & l_0  \frac{p_0 + p_1 + 2\beta}{2} + l_1  \frac{p_0 + p_1 - 2\beta}{2} = E_4 , \\
        & l_0  \frac{p_0 + p_1 - 2\alpha}{2} + l_1  \frac{p_0 + p_1 + 2\alpha}{2} = E_5 , \\
        & l_0  \frac{p_0 + p_1 + 2\alpha}{2} + l_1  \frac{p_0 + p_1 - 2\alpha}{2} = E_6 .
    \end{aligned}
    \right.
    \label{eq:S-tomo-system}
\end{equation}
Solving the linear system,
\begin{equation}
    \left\{
    \begin{aligned}
        & p_0 = \frac{1}{2} + \frac{E_1 - E_2}{2 L_{01}} , \\
        & p_1 = 1 - p_0 , \\
        & \alpha = \frac{E_6 - E_5}{2 L_{01}} , \\
        & \beta = \frac{E_4 - E_3}{2 L_{01}} ,
    \end{aligned}
    \right.
    \label{eq:S-tomo-solution}
\end{equation}
where the contrast $L_{01} = l_0 - l_1$ is calibrated from a Rabi oscillation experiment combined with the polarization $p_{\mathrm{e}} \approx 0.96$.
The imperfect polarization renders the measured state mixed.
Under any unitary evolution the mixed and pure components evolve independently,
\begin{equation}
    \rho_{\mathrm{mix}}
    = U \begin{pmatrix} p_{\mathrm{e}} & 0 \\ 0 & 1 - p_{\mathrm{e}} \end{pmatrix} U^\dagger
    = (2 p_{\mathrm{e}} - 1)  \rho_{\mathrm{pure}} + (1 - p_{\mathrm{e}})  I_2 ,
    \label{eq:S-mixed-state}
\end{equation}
with $I_2$ the qubit identity and $\rho_{\mathrm{pure}} = U |0\rangle\langle 0| U^\dagger$ the target pure state, so inverting Eq.~\eqref{eq:S-mixed-state} reconstructs the pure trajectory $\{|\psi(t)\rangle\}$ that enters the trace-distance analysis.
The uncertainties of the PL rates are dominated by photon shot noise; they propagate through Eq.~\eqref{eq:S-tomo-solution} to the density-matrix elements, and from there to the trace distances by Monte Carlo resampling of the reconstructed states, which defines the error bars and error bands quoted in the main text.

\subsection{Trace-distance evaluation and plateau analysis}
\label{sec:S4D}

From the reconstructed trajectory, the temporal moments are evaluated as discrete sums,
\begin{equation}
    \rho_T^{(k)} = \frac{1}{T} \sum_{t=0}^{T-1} \big(|\psi(t)\rangle\langle\psi(t)|\big)^{\otimes k} ,
    \qquad k = 1, \ldots, 4 ,
    \label{eq:S-discrete-moment}
\end{equation}
and compared with the Haar moments $\rho_{\mathrm{Haar}}^{(k)} = \Pi_{\mathrm{sym}} / (k+1)$ on the $(k{+}1)$-dimensional symmetric subspace, on which the trace distance $\Delta_T^{(k)}$ is evaluated.
Figure~4(a) of the main text displays the four orders for the two initial states at logarithmically spaced values of $T$: the first three trace distances decay without saturation, while $\Delta_T^{(4)}$ settles onto a nonzero plateau, at a higher value for $|\alpha\rangle$ than for $|0\rangle$.

The plateau values indicated by the dashed lines in Fig.~4(a) of the main text are obtained by averaging the experimental $\Delta_T^{(4)}$ over the window $T\in[1500,2000]$.
The resulting values are $0.00779$ for $|0\rangle$ and $0.00812$ for $|\alpha\rangle$.
Both values lie slightly above the asymptotic predictions, $1/160=0.00625$ and $0.00686$, respectively.
Part of this discrepancy arises from the finite observation window, since $\Delta_T^{(4)}$ has not yet fully reached its $T\to\infty$ limit.
To estimate this finite-window correction, we average the exact finite-$T$ theory over the same interval $T\in[1500,2000]$, obtaining $0.00650$ for $|0\rangle$ and $0.00707$ for $|\alpha\rangle$.
The remaining upward shift, of order $10^{-3}$ for both initial states, is attributed to experimental imperfections, including decoherence and control errors.
Importantly, these effects shift the absolute plateau values without changing their ordering: both the finite-$T$ theory and the experiment give a larger fourth-order deviation for $|\alpha\rangle$ than for $|0\rangle$, with the experimental central values preserving this ordering throughout the sampled range shown in Fig.~4(b).

\makeatletter
\begingroup
\let\@FMN@list\@empty
\let\label\@gobble
\makeatother
\endgroup

\end{document}